\documentclass[iop,revtex4-1]{hackemulateapj}
\usepackage{natbib}
\usepackage{hyperref}

\usepackage{amsmath}
\usepackage{xspace}
\usepackage{lineno}

\newcommand{\eg}{e.g.\xspace}

\mathchardef\mhyphen="2D
\newcommand{\vect}[1]{\boldsymbol{#1}}
\newcommand{\roughly}{\ensuremath{ {\sim}\,} }

\newlength{\dhatheight}

\newcommand{\code}[1]{\texttt{#1}\xspace}

\newcommand{\SExtractor}{\code{SExtractor}}

\newcommand{\spreadmodel}{\ensuremath{spread\_model}\xspace}

\newcommand{\age}{{\ensuremath{\tau}}\xspace}
\newcommand{\metal}{{\ensuremath{Z}}\xspace}

\newcommand{\unit}[1]{\ensuremath{\mathrm{\,#1}}\xspace}
\newcommand{\Gyr}{\unit{Gyr}}

\newcommand{\GeV}{\unit{GeV}}

\newcommand{\angstrom}{\unit{\AA}}

\newcommand{\cm}{\unit{cm}}
\newcommand{\km}{\unit{km}}
\newcommand{\pc}{\unit{pc}}
\newcommand{\kpc}{\unit{kpc}}
\newcommand{\Mpc}{\unit{Mpc}}
\newcommand{\second}{\unit{s}}

\newcommand{\Msun}{\unit{M_\odot}}

\newcommand{\Lsun}{\unit{L_\odot}}

\newcommand{\magn}{\unit{mag}}
\newcommand{\GeVcm}{\GeV^2 \cm^{-5}}
\newcommand{\kmss}{\km \second^{-1}}

\newcommand{\Jfactor}{J-factor\xspace}
\newcommand{\Jfactors}{J-factors\xspace}

\providecommand\physrep{\ref@jnl{Phys.~Rep.}}%
\providecommand\apjs{\ref@jnl{ApJS}}%
\providecommand{\jcap}{\ref@jnl{JCAP}}%
\newcommand{\kms}         {\ensuremath{\km \second^{-1}}\xspace}

\def\spose#1{\hbox to 0pt{#1\hss}}
\def\lta{\mathrel{\spose{\lower 3pt\hbox{$\mathchar"218$}}
     \raise 2.0pt\hbox{$\mathchar"13C$}}}
\def\gta{\mathrel{\spose{\lower 3pt\hbox{$\mathchar"218$}}
    \raise 2.0pt\hbox{$\mathchar"13E$}}}

\newcommand{\retII}{{Ret~II}\xspace}

\newcommand*\ruleline[1]{\par\noindent\raisebox{1ex}{\makebox[0.97\linewidth]{\hrulefill\quad\raisebox{-.6ex}{#1}\quad\hrulefill}}}

\newcommand{\vbulk}{\ensuremath{62.8 \pm 0.5 \kmss}\xspace}
\newcommand{\vdisp}{\ensuremath{3.3 \pm 0.7 \kmss}\xspace}
\newcommand{\mass}{\ensuremath{5.6 \pm 2.4 \times 10^{5} \Msun}\xspace}
\newcommand{\masstolight}{\ensuremath{470 \pm 210 \Msun/\Lsun}\xspace}
\newcommand{\feh}{\ensuremath{-2.65 \pm 0.07}\xspace}
\newcommand{\jsmall}{\ensuremath{18.8 \pm 0.6 \GeVcm}\xspace}
\newcommand{\jlarge}{\ensuremath{18.9 \pm 0.6 \GeVcm}\xspace}
\newcommand{\rhalf}{\ensuremath{R_{1/2}}\xspace}

\shorttitle{Kinematics and Metallicities in Reticulum II}
\shortauthors{Simon et al.}
\slugcomment{Accepted for publication in ApJ}

\begin{document}
%\linenumbers

\title{Stellar Kinematics and Metallicities in the Ultra-Faint Dwarf
  Galaxy Reticulum II\altaffilmark{*}\altaffilmark{\dag}\altaffilmark{\ddag}}

\altaffiltext{*}{This paper includes data gathered with the 6.5 meter
  Magellan Telescopes located at Las Campanas Observatory, Chile.}
 
\altaffiltext{\dag}{Based on observations obtained at the Gemini
  Observatory, which is operated by the Association of Universities
  for Research in Astronomy, Inc., under a cooperative agreement with
  the NSF on behalf of the Gemini partnership: the National Science
  Foundation (United States), the National Research Council (Canada),
  CONICYT (Chile), the Australian Research Council (Australia),
  Minist\'{e}rio da Ci\^{e}ncia, Tecnologia e Inova\c{c}\~{a}o
  (Brazil) and Ministerio de Ciencia, Tecnolog\'{i}a e Innovaci\'{o}n
  Productiva (Argentina).}
 
\altaffiltext{\ddag}{Based on data obtained from the ESO Science
  Archive Facility under request number 157689.}

%\author{Joshua D. Simon and many others}
%\altaffiltext{1}{Observatories of the Carnegie Institution of
%  Washington, 813 Santa Barbara St., Pasadena, CA 91101;
%  jsimon@obs.carnegiescience.edu}

%\title{Stellar Kinematics and Metallicities in the Ultra-Faint Dwarf
%  Galaxy Reticulum II}

% WARNING: Hacking umlauts for Munich affiliation...

\author{J.~D.~Simon}
\affiliation{Carnegie Observatories, 813 Santa Barbara St., Pasadena, CA 91101, USA}
\author{A.~Drlica-Wagner}
\affiliation{Fermi National Accelerator Laboratory, P. O. Box 500, Batavia, IL 60510, USA}
\author{T.~S.~Li}
\affiliation{George P. and Cynthia Woods Mitchell Institute for Fundamental Physics and Astronomy, and Department of Physics and Astronomy, Texas A\&M University, College Station, TX 77843,  USA}
\author{B.~Nord}
\affiliation{Fermi National Accelerator Laboratory, P. O. Box 500, Batavia, IL 60510, USA}
\author{M.~Geha}
\affiliation{Astronomy Department, Yale University, New Haven, CT 06520, USA}
\author{K.~Bechtol}
\affiliation{Kavli Institute for Cosmological Physics, University of Chicago, Chicago, IL 60637, USA}
\author{E.~Balbinot}
\affiliation{Department of Physics, University of Surrey, Guildford GU2 7XH, UK}
\affiliation{Laborat\'orio Interinstitucional de e-Astronomia - LIneA, Rua Gal. Jos\'e Cristino 77, Rio de Janeiro, RJ - 20921-400, Brazil}
\author{E.~Buckley-Geer}
\affiliation{Fermi National Accelerator Laboratory, P. O. Box 500, Batavia, IL 60510, USA}
\author{H.~Lin}
\affiliation{Fermi National Accelerator Laboratory, P. O. Box 500, Batavia, IL 60510, USA}
\author{J.~Marshall}
\affiliation{George P. and Cynthia Woods Mitchell Institute for Fundamental Physics and Astronomy, and Department of Physics and Astronomy, Texas A\&M University, College Station, TX 77843,  USA}
\author{B.~Santiago}
\affiliation{Instituto de F\'\i sica, UFRGS, Caixa Postal 15051, Porto Alegre, RS - 91501-970, Brazil}
\affiliation{Laborat\'orio Interinstitucional de e-Astronomia - LIneA, Rua Gal. Jos\'e Cristino 77, Rio de Janeiro, RJ - 20921-400, Brazil}
\author{L.~Strigari}
\affiliation{George P. and Cynthia Woods Mitchell Institute for Fundamental Physics and Astronomy, and Department of Physics and Astronomy, Texas A\&M University, College Station, TX 77843,  USA}
\author{M.~Wang}
\affiliation{George P. and Cynthia Woods Mitchell Institute for Fundamental Physics and Astronomy, and Department of Physics and Astronomy, Texas A\&M University, College Station, TX 77843,  USA}
\author{R.~H.~Wechsler}
\affiliation{Department of Physics, Stanford University, 382 Via Pueblo Mall, Stanford, CA 94305, USA}
\affiliation{Kavli Institute for Particle Astrophysics \& Cosmology, P. O. Box 2450, Stanford University, Stanford, CA 94305, USA}
\affiliation{SLAC National Accelerator Laboratory, Menlo Park, CA 94025, USA}
\author{B.~Yanny}
\affiliation{Fermi National Accelerator Laboratory, P. O. Box 500, Batavia, IL 60510, USA}
\author{T.~Abbott}
\affiliation{Cerro Tololo Inter-American Observatory, National Optical Astronomy Observatory, Casilla 603, La Serena, Chile}
\author{A.H.~Bauer}
\affiliation{Institut de Ci\`encies de l'Espai, IEEC-CSIC, Campus UAB, Facultat de Ci\`encies, Torre C5 par-2, 08193 Bellaterra, Barcelona, Spain}
\author{G.~M.~Bernstein}
\affiliation{Department of Physics and Astronomy, University of Pennsylvania, Philadelphia, PA 19104, USA}
\author{E.~Bertin}
\affiliation{Sorbonne Universit\'es, UPMC Univ Paris 06, UMR 7095, Institut d'Astrophysique de Paris, F-75014, Paris, France}
\affiliation{Institut d'Astrophysique de Paris, Univ. Pierre et Marie Curie \& CNRS UMR7095, F-75014 Paris, France}
\author{D.~Brooks}
\affiliation{Department of Physics \& Astronomy, University College London, Gower Street, London, WC1E 6BT, UK}
\author{D.~L.~Burke}
\affiliation{Kavli Institute for Particle Astrophysics \& Cosmology, P. O. Box 2450, Stanford University, Stanford, CA 94305, USA}
\affiliation{SLAC National Accelerator Laboratory, Menlo Park, CA 94025, USA}
\author{D.~Capozzi}
\affiliation{Institute of Cosmology \& Gravitation, University of Portsmouth, Portsmouth, PO1 3FX, UK}
\author{A.~Carnero~Rosell}
\affiliation{Laborat\'orio Interinstitucional de e-Astronomia - LIneA, Rua Gal. Jos\'e Cristino 77, Rio de Janeiro, RJ - 20921-400, Brazil}
\affiliation{Observat\'orio Nacional, Rua Gal. Jos\'e Cristino 77, Rio de Janeiro, RJ - 20921-400, Brazil}
\author{M.~Carrasco~Kind}
\affiliation{Department of Astronomy, University of Illinois, 1002 W. Green Street, Urbana, IL 61801, USA}
\affiliation{National Center for Supercomputing Applications, 1205 West Clark St., Urbana, IL 61801, USA}
\author{C.~B.~D'Andrea}
\affiliation{Institute of Cosmology \& Gravitation, University of Portsmouth, Portsmouth, PO1 3FX, UK}
\author{L.~N.~da Costa}
\affiliation{Laborat\'orio Interinstitucional de e-Astronomia - LIneA, Rua Gal. Jos\'e Cristino 77, Rio de Janeiro, RJ - 20921-400, Brazil}
\affiliation{Observat\'orio Nacional, Rua Gal. Jos\'e Cristino 77, Rio de Janeiro, RJ - 20921-400, Brazil}
\author{D.~L.~DePoy}
\affiliation{George P. and Cynthia Woods Mitchell Institute for Fundamental Physics and Astronomy, and Department of Physics and Astronomy, Texas A\&M University, College Station, TX 77843,  USA}
\author{S.~Desai}
\affiliation{Department of Physics, Ludwig-Maximilians-Universit\"at, Scheinerstr.\ 1, 81679 M\"unchen, Germany}
\author{H.~T.~Diehl}
\affiliation{Fermi National Accelerator Laboratory, P. O. Box 500, Batavia, IL 60510, USA}
\author{S.~Dodelson}
\affiliation{Fermi National Accelerator Laboratory, P. O. Box 500, Batavia, IL 60510, USA}
\affiliation{Kavli Institute for Cosmological Physics, University of Chicago, Chicago, IL 60637, USA}
\author{C.~E Cunha}
\affiliation{Kavli Institute for Particle Astrophysics \& Cosmology, P. O. Box 2450, Stanford University, Stanford, CA 94305, USA}
\author{J.~Estrada}
\affiliation{Fermi National Accelerator Laboratory, P. O. Box 500, Batavia, IL 60510, USA}
\author{A.~E.~Evrard}
\affiliation{Department of Physics, University of Michigan, Ann Arbor, MI 48109, USA}
\author{A.~Fausti Neto}
\affiliation{Laborat\'orio Interinstitucional de e-Astronomia - LIneA, Rua Gal. Jos\'e Cristino 77, Rio de Janeiro, RJ - 20921-400, Brazil}
\author{E.~Fernandez}
\affiliation{Institut de F\'{\i}sica d'Altes Energies, Universitat Aut\`onoma de Barcelona, E-08193 Bellaterra, Barcelona, Spain}
\author{D.~A.~Finley}
\affiliation{Fermi National Accelerator Laboratory, P. O. Box 500, Batavia, IL 60510, USA}
\author{B.~Flaugher}
\affiliation{Fermi National Accelerator Laboratory, P. O. Box 500, Batavia, IL 60510, USA}
\author{J.~Frieman}
\affiliation{Fermi National Accelerator Laboratory, P. O. Box 500, Batavia, IL 60510, USA}
\affiliation{Kavli Institute for Cosmological Physics, University of Chicago, Chicago, IL 60637, USA}
\author{E.~Gaztanaga}
\affiliation{Institut de Ci\`encies de l'Espai, IEEC-CSIC, Campus UAB, Facultat de Ci\`encies, Torre C5 par-2, 08193 Bellaterra, Barcelona, Spain}
\author{D.~Gerdes}
\affiliation{Department of Physics, University of Michigan, Ann Arbor, MI 48109, USA}
\author{D.~Gruen}
\affiliation{Max Planck Institute for Extraterrestrial Physics, Giessenbachstrasse, 85748 Garching, Germany}
\affiliation{University Observatory Munich, Scheinerstrasse 1, 81679 Munich, Germany}
\author{R.~A.~Gruendl}
\affiliation{Department of Astronomy, University of Illinois, 1002 W. Green Street, Urbana, IL 61801, USA}
\affiliation{National Center for Supercomputing Applications, 1205 West Clark St., Urbana, IL 61801, USA}
\author{K.~Honscheid}
\affiliation{Center for Cosmology and Astro-Particle Physics, The Ohio State University, Columbus, OH 43210, USA}
\affiliation{Department of Physics, The Ohio State University, Columbus, OH 43210, USA}
\author{D.~James}
\affiliation{Cerro Tololo Inter-American Observatory, National Optical Astronomy Observatory, Casilla 603, La Serena, Chile}
\author{S.~Kent}
\affiliation{Fermi National Accelerator Laboratory, P. O. Box 500, Batavia, IL 60510, USA}
\author{K.~Kuehn}
\affiliation{Australian Astronomical Observatory, North Ryde, NSW 2113, Australia}
\author{N.~Kuropatkin}
\affiliation{Fermi National Accelerator Laboratory, P. O. Box 500, Batavia, IL 60510, USA}
\author{O.~Lahav}
\affiliation{Department of Physics \& Astronomy, University College London, Gower Street, London, WC1E 6BT, UK}
\author{M.~A.~G.~Maia}
\affiliation{Laborat\'orio Interinstitucional de e-Astronomia - LIneA, Rua Gal. Jos\'e Cristino 77, Rio de Janeiro, RJ - 20921-400, Brazil}
\affiliation{Observat\'orio Nacional, Rua Gal. Jos\'e Cristino 77, Rio de Janeiro, RJ - 20921-400, Brazil}
\author{M.~March}
\affiliation{Department of Physics and Astronomy, University of Pennsylvania, Philadelphia, PA 19104, USA}
\author{P.~Martini}
\affiliation{Center for Cosmology and Astro-Particle Physics, The Ohio State University, Columbus, OH 43210, USA}
\affiliation{Department of Astronomy, The Ohio State University, Columbus, OH 43210, USA}
\author{C.~J.~Miller}
\affiliation{Department of Astronomy, University of Michigan, Ann Arbor, MI 48109, USA}
\affiliation{Department of Physics, University of Michigan, Ann Arbor, MI 48109, USA}
\author{R.~Miquel}
\affiliation{Institut de F\'{\i}sica d'Altes Energies, Universitat Aut\`onoma de Barcelona, E-08193 Bellaterra, Barcelona, Spain}
\author{R.~Ogando}
\affiliation{Laborat\'orio Interinstitucional de e-Astronomia - LIneA, Rua Gal. Jos\'e Cristino 77, Rio de Janeiro, RJ - 20921-400, Brazil}
\affiliation{Observat\'orio Nacional, Rua Gal. Jos\'e Cristino 77, Rio de Janeiro, RJ - 20921-400, Brazil}
\author{A.~K.~Romer}
\affiliation{Department of Physics and Astronomy, Pevensey Building, University of Sussex, Brighton, BN1 9QH, UK}
\author{A.~Roodman}
\affiliation{Kavli Institute for Particle Astrophysics \& Cosmology, P. O. Box 2450, Stanford University, Stanford, CA 94305, USA}
\affiliation{SLAC National Accelerator Laboratory, Menlo Park, CA 94025, USA}
\author{E.~S.~Rykoff}
\affiliation{Kavli Institute for Particle Astrophysics \& Cosmology, P. O. Box 2450, Stanford University, Stanford, CA 94305, USA}
\affiliation{SLAC National Accelerator Laboratory, Menlo Park, CA 94025, USA}
\author{M.~Sako}
\affiliation{Department of Physics and Astronomy, University of Pennsylvania, Philadelphia, PA 19104, USA}
\author{E.~Sanchez}
\affiliation{Centro de Investigaciones Energ\'eticas, Medioambientales y Tecnol\'ogicas (CIEMAT), Madrid, Spain}
\author{M.~Schubnell}
\affiliation{Department of Physics, University of Michigan, Ann Arbor, MI 48109, USA}
\author{I.~Sevilla}
\affiliation{Centro de Investigaciones Energ\'eticas, Medioambientales y Tecnol\'ogicas (CIEMAT), Madrid, Spain}
\affiliation{Department of Astronomy, University of Illinois, 1002 W. Green Street, Urbana, IL 61801, USA}
\author{R.~C.~Smith}
\affiliation{Cerro Tololo Inter-American Observatory, National Optical Astronomy Observatory, Casilla 603, La Serena, Chile}
\author{M.~Soares-Santos}
\affiliation{Fermi National Accelerator Laboratory, P. O. Box 500, Batavia, IL 60510, USA}
\author{F.~Sobreira}
\affiliation{Fermi National Accelerator Laboratory, P. O. Box 500, Batavia, IL 60510, USA}
\affiliation{Laborat\'orio Interinstitucional de e-Astronomia - LIneA, Rua Gal. Jos\'e Cristino 77, Rio de Janeiro, RJ - 20921-400, Brazil}
\author{E.~Suchyta}
\affiliation{Center for Cosmology and Astro-Particle Physics, The Ohio State University, Columbus, OH 43210, USA}
\affiliation{Department of Physics, The Ohio State University, Columbus, OH 43210, USA}
\author{M.~E.~C.~Swanson}
\affiliation{National Center for Supercomputing Applications, 1205 West Clark St., Urbana, IL 61801, USA}
\author{G.~Tarle}
\affiliation{Department of Physics, University of Michigan, Ann Arbor, MI 48109, USA}
\author{J.~Thaler}
\affiliation{Department of Physics, University of Illinois, 1110 W. Green St., Urbana, IL 61801, USA}
\author{D.~Tucker}
\affiliation{Fermi National Accelerator Laboratory, P. O. Box 500, Batavia, IL 60510, USA}
\author{V.~Vikram}
\affiliation{Argonne National Laboratory, 9700 South Cass Avenue, Lemont, IL 60439, USA}
\author{A.~R.~Walker}
\affiliation{Cerro Tololo Inter-American Observatory, National Optical Astronomy Observatory, Casilla 603, La Serena, Chile}
\author{W.~Wester}
\affiliation{Fermi National Accelerator Laboratory, P. O. Box 500, Batavia, IL 60510, USA}

\collaboration{The DES Collaboration}

\begin{abstract}
We present Magellan/M2FS, VLT/GIRAFFE, and Gemini South/GMOS
spectroscopy of the newly discovered Milky Way satellite Reticulum II.
Based on the spectra of 25 \retII member stars selected from Dark
Energy Survey imaging, we measure a mean heliocentric velocity of
\vbulk and a velocity dispersion of \vdisp.  The mass-to-light ratio
of \retII within its half-light radius is \masstolight, demonstrating
that it is a strongly dark matter-dominated system.  Despite its
spatial proximity to the Magellanic Clouds, the radial velocity of
\retII differs from that of the LMC and SMC by 199 and 83~\kms,
respectively, suggesting that it is not gravitationally bound to the
Magellanic system.  The likely member stars of \retII span 1.3~dex in
metallicity, with a dispersion of $0.28 \pm 0.09$~dex, and we identify
several extremely metal-poor stars with $\mbox{[Fe/H]} < -3$.  In
combination with its luminosity, size, and ellipticity, these results
confirm that \retII is an ultra-faint dwarf galaxy.  With a mean
metallicity of $\mbox{[Fe/H]} = \feh$, \retII matches Segue~1 as the
most metal-poor galaxy known.  Although \retII is the third-closest
dwarf galaxy to the Milky Way, the line-of-sight integral of the dark
matter density squared is $\log_{10}(J) = \jsmall$ within $0.2\degr$,
indicating that the predicted gamma-ray flux from dark matter
annihilation in \retII is lower than that of several other dwarf
galaxies.
\end{abstract}


\section{INTRODUCTION}
\label{intro}

The population of known dwarf galaxies orbiting the Milky Way has
grown rapidly over the last decade, with the discovery of the
ultra-faint dwarfs by the Sloan Digital Sky Survey (SDSS) more than
doubling the size of our Galaxy's retinue of satellites
\citep[e.g.,][]{willman05,zucker06,belokurov07}.  These extreme
objects are the focus of an enormous variety of ongoing work, ranging
from their internal kinematics
\citep[e.g.,][]{martin07,sg07,koposov11}, metallicities
\citep[e.g.,][]{kirby08,norris10}, and chemical abundance patterns
\citep[e.g.,][]{koch08,frebel10,vargas13,fsk14} to their star
formation histories \citep[e.g.,][]{weisz14a,brown14}, cosmological
implications \citep[e.g.,][]{weisz14b}, and ability to constrain dark
matter models via indirect detection
\citep[e.g.,][]{strigari07,Ackermann:2013yva,Geringer-Sameth:2014qqa}.
However, progress has slowed in the past five years as the flow of
discoveries from SDSS dwindled (although see \citealt{laevens15},
\citealt{martin15}, and \citealt{kim15}).

The recent discovery of eight new candidate dwarf galaxies
\citep{bechtol15,koposov15} in $\roughly 2000 \deg^{2}$ of DECam imaging
data from the first year of the Dark Energy Survey
\citep[DES;][]{decam,diehl14} promises to reinvigorate studies of the
faintest galaxies.  Some of the particularly interesting aspects of
these newly discovered objects are their apparent concentration around
the Magellanic Clouds, the identification of a relatively luminous
dwarf (Eridanus~II) near the Milky Way's virial radius with possible
recent star formation, and the existence of hyper-faint ($M_{V}
\lesssim -3$) dwarfs beyond the immediate vicinity of the Milky Way.
Four of these new satellites --- Reticulum~II, Eridanus~II, Tucana~II,
and Horologium~I --- can be fairly confidently classified as galaxies
based on DES imaging alone, while the other systems have physical
sizes and luminosities that overlap within their uncertainties with
those of some globular clusters.

In this paper we begin the process of spectroscopic follow-up
observations of the new DES ultra-faint satellites.  We present low
and high resolution spectroscopy of stars in \retII, the closest
($d=32$~kpc) and best-characterized of the DES satellites, with an
ellipticity of 0.6 and a projected elliptical half-light radius of
0.1\degr~\citep{bechtol15}.  The velocities and metallicities of these
stars confirm that \retII is indeed a dwarf galaxy.  In
\S\ref{observations} we describe our spectroscopic target selection,
observations, and data reduction.  We discuss the measurement of
stellar velocities and metallicities and the classification of \retII
member stars in \S\ref{measurements}.  We focus on some of the
implications of our results and examine how \retII fits into the
previously known population of Milky Way satellite galaxies in
\S\ref{discussion}.  In \S\ref{conclusions} we summarize our findings
and conclude.

\section{OBSERVATIONS AND DATA REDUCTION}
\label{observations}

\subsection{Target Selection}
\label{sec:targets}

Spectroscopic targets were selected from the object catalog derived
from the coadded images of the first internal annual release of DES
data \citep[Y1A1; Gruendl et al., in
  prep;][]{Sevilla:2011,2012ApJ...757...83D,2012SPIE.8451E..0DM,balbinot15}.
We identified objects as stars based on the \spreadmodel quantity
output by \SExtractor~\citep{2011ASPC..442..435B,2012ApJ...757...83D}.
Our stellar sample consists of well-measured objects with
$|spread\_model\_i| \allowbreak < \allowbreak 0.002$ and
$flags\_\{g,r,i\} < 4$.  We selected likely members based on a
matched-filter maximum likelihood procedure combining a spatial model
of \retII and the predicted color-magnitude distribution of an old
metal poor stellar population~\citep{bechtol15}.  We modeled the
spatial distribution of \retII with the best-fit elliptical Plummer
profile with ellipticity of 0.6 and an elliptical half-light radius of
$0.1\degr$.  The distribution of stars in color-magnitude space was
modeled by a composite of four isochrones bracketing a range of ages,
$\age = \{ 12.6 \Gyr, 14.1 \Gyr \}$, and metallicities, $\metal = \{
0.0001, 0.0002
\}$~\citep{2008A&A...482..883M,2010ApJ...724.1030G}. These isochrones
are placed at the best-fit distance modulus of $m-M = 17.5$ and
weighted by the initial mass function of \citet{2001ApJ...554.1274C}.
We used a composite isochrone to maximize the number of candidate
members while remaining agnostic to the exact age and metallicity of
\retII.

One result of the maximum-likelihood fit is a membership probability
for each object in our stellar
catalog~\citep{bechtol15,2009ApJ...703..601R}.  This membership
probability incorporates both the spatial separation of the star from
the centroid of \retII and the distance of the object from the
composite isochrone in color-magnitude space.  Our spectroscopic
follow-up sample primarily consisted of stars with membership
probability $p > 0.01$ within $0.5\degr$ of the centroid of \retII.
Each of these objects was visually inspected in the DES coadded images and
the imaging was confirmed to be of high quality.

We targeted \retII for follow-up spectroscopy with both Magellan/M2FS
and Gemini/GMOS, and we have also made use of publicly available
VLT/GIRAFFE spectra of \retII stars in the ESO Archive.  For
observations with the 256-fiber Magellan/M2FS spectrograph
(\S\ref{m2fs}), we targeted objects within $14.65\arcmin$ of \retII
with $p \geq 0.01$ and $g \leq 22$.  A small fraction of the 177
targets meeting these criteria were not observed because of fiber
collisions and exclusion regions around bright stars.  Given the
availability of additional fibers, we also added 18 stars located near
the \retII isochrone that missed the probability and/or photometric
quality cuts.  Out of these 195 stars, a total of 185 were targeted
with science fibers.  Six fibers were broken, and the remaining 65
were placed on blank sky positions chosen to have low count rates from
a Digitized Sky Survey image.

The Gemini/GMOS observation field (\S\ref{gmos}) was selected as the
$5\arcmin \times 5\arcmin$ region with the highest summed membership
probability.  Slits were iteratively prioritized based on the
membership probability and brightness of targets, focusing on stars on
the giant branch for which metallicities could be measured via the Ca
triplet (CaT) lines.  Final mask creation was performed using the GMMPS
v0.402 toolkit, which was used to assign slits to 33 science targets
and 4 acquisition
objects.\footnote{\url{http://www.gemini.edu/node/12255?q=node/10458}}

VLT/GIRAFFE observations of \retII (\S\ref{vlt}) were carried out under
the auspices of the Gaia/ESO Survey \citep{gilmore12}, and the target
selection was performed by the Gaia/ESO Survey team independently of
our DES photometric analysis chain.  It is clear from the colors and
magnitudes of the observed stars that they were chosen to lie near the
\retII fiducial sequence determined by \citet{koposov15}, but we do
not have access to the specific criteria by which priorities were set.
Comparing this sample with our catalog, we find that 14 of the stars
observed are high-probability member candidates ($p > 0.5$), 8 have
low to intermediate membership probabilities ($0.5 > p > 0.01$), and
23 are unlikely to be members ($p < 0.01$).  The majority (62) of the
VLT spectra are of bright stars not included in the DES coadd catalog
because of its saturation limit at $g \sim 17$, but (with one
exception; \S\ref{giraffe_gmos_mem}) these stars appear to have been
selected to be Milky Way disk stars and are not relevant for our
purposes.

\subsection{Magellan/M2FS Spectra}
\label{m2fs}

We observed \retII with the multi-fiber M2FS spectrograph
\citep{m2fs} on the Magellan/Clay Telescope on 2015 February 19.  M2FS
consists of two identical spectrographs, generally referred to as
``red'' and ``blue'' even though neither is optimized for a particular
wavelength range.  Each spectrograph is fed by 128 1.2\arcsec\ fibers,
which are positioned on the sky over a 29.5\arcmin\ diameter field
with a plug plate.  We used M2FS in its high resolution configuration
at $R \approx 25000$, with a narrow-band filter to isolate a single
spectral order covering the Mg~b spectral region ($5120-5190$~\AA) for
each fiber.  The 256 M2FS fibers were placed on 185 stars selected
from the DES Y1A1 internal data release, as well as 65 blank sky
positions for sky subtraction, as described in \S\ref{sec:targets}.

We obtained three $2400\second$ exposures on the \retII field in
variable seeing conditions and with decreased transparency resulting
from thin clouds.  We also obtained twilight sky spectra for
flatfielding, ThArNe lamp frames for wavelength calibration, and
spectra of the K giant radial velocity standard star CD$-43\degr$2527
through a single fiber to check the velocity zero point.

We reduced the data with the \code{dohydra} package in IRAF.  We began
by subtracting the bias level, recombining the files from the four
amplifiers used to read out each CCD into a single frame, and masking
out cosmic rays using the L.A.Cosmic routine \citep{vandokkum01}.  We
then trimmed the frames to the central $\sim1000$ pixels to eliminate
contamination from neighboring spectral orders.  We traced the fiber
positions as a function of wavelength using a twilight frame and
removed scattered light with a two-dimensional fit to the areas
between each group of 16 fibers.  Because of the low signal-to-noise
ratio (S/N) of the data (the brightest stars reach $\rm{S/N} \approx
20$~pixel$^{-1}$, and the large majority of the targets are at
$\rm{S/N} < 10$~pixel$^{-1}$) and the excellent flatness of the e2V
detectors ($\sim1$~\%), we did not attempt to apply a flatfield
correction \citep{mh13}.  We extracted the spectrum of each fiber in
the ThArNe lamp, twilight, standard star, and science frames.  We also
performed the same extraction on a variance image derived from the
science frames to provide spectra of the pixel uncertainties.  We
determined the wavelength solution by fitting a fourth order
polynomial to known Th and Ar line wavelengths on the ThArNe
exposures.  The positions of the ThAr lines shifted by one pixel
(3.8~\kms) in the wavelength direction between the ThArNe frame taken
immediately before \retII was observed and the one taken immediately
after.  By examining the absorption lines of the highest S/N stars, we
determined that in the blue spectrograph, this shift occurred between
the second and third \retII exposures.  In the red spectrograph all
three \retII frames appeared to be aligned with each other.  We
therefore shifted the third frame taken with the blue spectrograph by
one pixel and re-ran the reductions using only the ThArNe frame from
before the \retII observing sequence (to maintain consistency between
the velocity scales for the red and blue spectrographs, the wavelength
solution for the red data was also determined with the ThArNe frame
observed before \retII).  This wavelength solution was then applied to
the data and the spectra were rebinned.  We constructed master sky
spectra for the red and blue spectrographs by coadding the $\sim30$
sky fiber spectra obtained in each spectrograph.  The sky spectra were
scaled according to the throughput of each target fiber (determined
from the twilight sky frames) and subtracted.

\subsection{Gemini/GMOS Spectra}
\label{gmos}

We also observed \retII with the GMOS-S spectrograph \citep{hook04} on
the Gemini South Telescope in queue mode on several nights beginning
on 2015 February 15 through program GS-2014B-DD-8.  We targeted 33
likely \retII member stars with a multi-slit mask, using
0.75\arcsec\ slits and a minimum slit length of 5\arcsec.  Our
observations employed the R831 grating to produce a spectral
resolution of $R=4400$ over a range of 2300\,\AA\ per spectrum, and
the RG610 filter to block second order light.  A total of eleven
$1200\second$ science-quality exposures were obtained.  We observed at
two different central wavelengths around the near-infrared CaT lines
(8550\,\AA\ and 8650\,\AA) to ensure that key spectral features did
not land in gaps between the three CCDs.  Science observations were
alternated with Quartz-Halogen flats and CuAr arc lamp exposures for
calibration.

We reduced the GMOS data using version 2.16 of the Gemini package in
IRAF.  For each dither on each night of observations, we created a
flat field frame and used it to process the science frame(s).  We
combined exposures from the same dithers and then derived the
wavelength solution.  The spectra were rectified and transformed onto
a common wavelength scale, and finally the dithers were coadded to
eliminate the CCD gaps.  We performed sky subtraction on each slit by
masking out the stellar continuum and fitting a linear function to the
background in the spatial direction.  Extractions were carried out on
the sky-subtracted spectra.

\subsection{VLT/GIRAFFE Spectra}
\label{vlt}

\retII was observed with the FLAMES/GIRAFFE spectrograph
\citep{pasquini00} on the VLT/UT2 telescope as part of the Gaia/ESO
survey \citep{gilmore12}.  Because the Gaia/ESO survey is a public
spectroscopic survey, these data are public and available from the ESO
Archive\footnote{http://archive.eso.org/cms/eso-data.html} immediately
after the observations are obtained.

Observations were taken in MEDUSA mode, which allows the simultaneous
observation of up to 132 objects.  On the night of 2015 March 9,
$2\times1500\second$ exposure were taken using the HR21 grating, which
covers the wavelength range from 8482--8981\angstrom at a resolution
of $R \sim 16200$.  The same field was observed with a bluer
wavelength setting in 2015 February, but because of the lower
signal-to-noise ratio we do not make use of those data in this paper.
The calibration frames taken as part of the Gaia/ESO Survey
observations include 5~biases, 3~flats, and 1~ThAr arc taken at the
end of the night.  We reduced the data with the GIRAFFE \code{Gasgano}
pipeline (v2.4.8) provided by
ESO.\footnote{http://www.eso.org/sci/software/gasgano.html} This
pipeline provides utilities for bias subtraction, flat-fielding,
spectral extraction of individual objects and accurate wavelength
calibration.

The CaT absorption lines are located close to a number of
bright sky emission lines, so we require accurate sky subtraction to
recover their positions and equivalent widths.  We first determined
fiber to fiber wavelength offsets for the 16 GIRAFFE sky fibers using
the night sky lines, finding that there are shifts of
$\sim0.15$~\kms between fibers.  We then shifted the sky spectra
based on these calculated offsets, linearly interpolated them, and
combined them into a master sky spectrum, rejecting outliers with a
$3\sigma$ clip.  For each science spectrum, we determined the
wavelength offset relative to the master sky and shifted the data
accordingly.  We scaled the sky spectrum to match the amplitude of the
bright, isolated sky lines at 8885.85~\AA\ and 8919.6~\AA\ and
subtracted it from the science spectrum.

\begin{figure*}[th!]
\epsscale{1.2}
\plotone{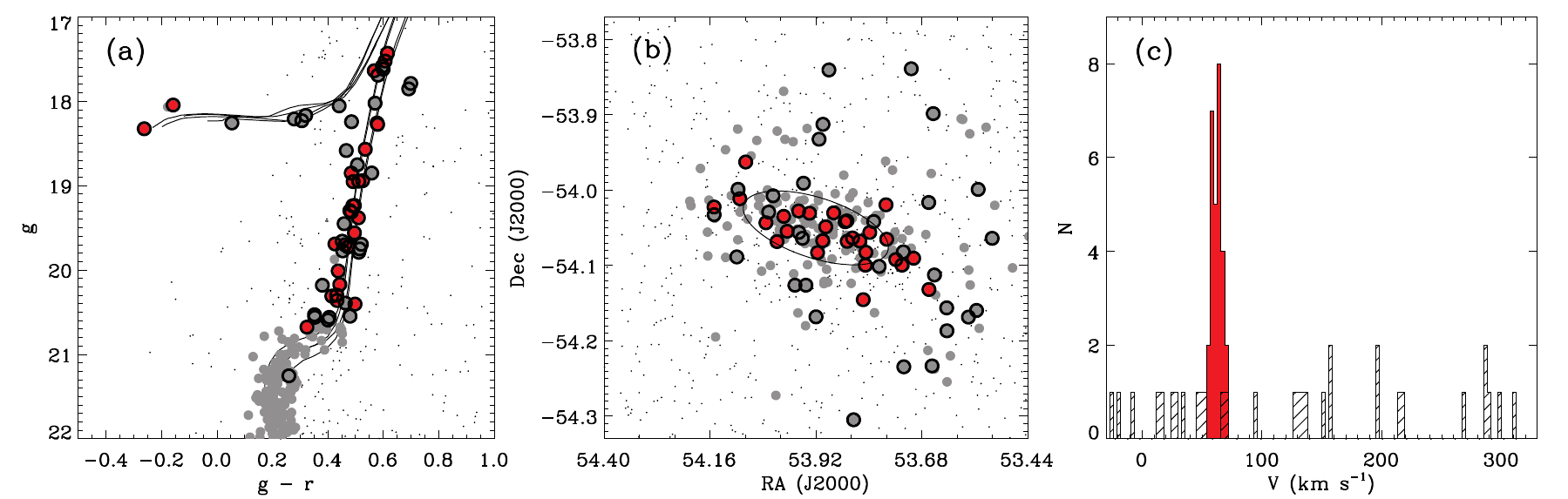}
\caption{\emph{(a)} DES color-magnitude diagram of Reticulum~II.
  Stars within 14.65\arcmin\ of the center of \retII are plotted as
  small black dots, and stars selected for spectroscopy with M2FS,
  GIRAFFE, and GMOS (as described in \S\ref{sec:targets}) are plotted
  as filled gray circles.  Points surrounded by black outlines
  represent the stars for which we obtained successful velocity
  measurements, and those we identify as \retII members are filled in
  with red.  The four PARSEC isochrones used to determine membership
  probabilities are displayed as black lines.  \emph{(b)} Spatial
  distribution of the observed stars.  Symbols are as in panel
  \emph{(a)}.  The half-light radius of \retII from \citet{bechtol15}
  is outlined as a black ellipse.  \emph{(c)} Radial velocity
  distribution of observed stars, combining all three spectroscopic
  data sets.  The clear narrow peak of stars at $v \sim 60$~\kms
  highlighted in red is the signature of \retII.  The hatched
  histogram indicates stars that are not members of \retII; note that
  there are two bins containing non-member stars near $v = 70$~\kms
  that are over-plotted on top of the red histogram..}
\label{cmd}
\end{figure*}

\section{VELOCITY AND METALLICITY MEASUREMENTS}
\label{measurements}

Because the M2FS spectra have higher spectral resolution than the
other data sets and cover the largest sample of candidate \retII
member stars, they are particularly well suited for measuring the
stellar kinematics of \retII.  The GIRAFFE spectra could also be used
for accurate radial velocity measurements \citep[e.g.,][]{koposov11},
but they include a smaller number of likely \retII members, nearly all
of which overlap with the M2FS target list.  In order to avoid the
complications of placing multiple instruments on a consistent velocity
scale, we choose to use M2FS alone for the primary kinematic results
reported in this paper.  However, as we note in \S\ref{sigma}, we
obtain consistent results with the GIRAFFE velocities.  The GMOS
observations were made at much lower spectral resolution and would not
be competitive with M2FS or GIRAFFE for velocity work.  GIRAFFE and
GMOS velocities for stars not observed with M2FS are listed in
Table~\ref{tab:spec} but are not used to constrain the velocity
dispersion of \retII.

The wavelength range covered by the M2FS observations is not
well-suited to metallicity measurements using the spectra alone,
because the strength of the Mg~b triplet lines depends on a number of
stellar properties: effective temperature, surface gravity,
metallicity, and Mg abundance.  No calibrations for the metallicities
of individual stars as a function of Mg~b strength are available in
the literature.  We therefore rely on the better-studied CaT
 region of the spectrum in the VLT and Gemini data for
metallicity information.

The primary data used in our analysis for the remainder of the paper
are 52 M2FS velocity measurements, 16 GIRAFFE metallicities, and 6
GMOS metallicities.  All velocity measurements presented here have
been transformed to the heliocentric frame.  We use heliocentric
corrections of $-8.26$~\kms\ for the M2FS observations and
$-5.63$~\kms\ for the GIRAFFE observations.

\subsection{Radial Velocity Measurements}

Given a velocity data set in which the intrinsic velocity dispersion
may not be much larger than the uncertainties on individual velocity
measurements, accurate characterization of both the velocities and the
measurement uncertainties is critical for a reliable determination of
the velocity dispersion.  We used Monte Carlo simulations to
investigate the statistical uncertainties of our velocity
measurements, and repeat observations to constrain systematic
uncertainties.

We determined radial velocities using the same basic approach as in
\citet{sg07} and subsequent papers.  We performed a $\chi^{2}$ fit to
each spectrum with a high S/N template spectrum as a function of the
velocity offset between the two \citep{lee08,sdssdr6,newman13}.  As a
template spectrum for this analysis, we chose a high resolution
spectrum of the metal-poor giant HD~122563 obtained in 2009 with
Magellan/MIKE, because the M2FS RV standard CD$-43\degr$2527 is
relatively metal-rich and does not provide a very good match to the
spectra of the \retII stars.  We assume a heliocentric velocity of
$-26.5$~\kms for HD~122563 \citep{chubak12}. We then ran 500 Monte
Carlo simulations per star, in which noise comparable to the
calculated variance in each pixel was added to the spectrum and the
velocity was re-measured as above. The statistical uncertainty on the
velocity of each star was defined to be the square root of the
variance in the measured mean velocity for the 500 Monte Carlo
spectra.  Highly discrepant Monte Carlo iterations were discarded
before computing the variance.  The median statistical uncertainty on
the velocity measurements for the M2FS data set is 1.0~\kms.

As a test of systematic velocity uncertainties in the M2FS data, we
reduced the three \retII science frames separately, and measured
velocities for the brightest stars in each frame.  Despite carrying
out these reductions in multiple ways and computing velocities with
two different techniques, we found that the velocity difference
between independent measurements of a star on frames $i$ and $j$,
$\Delta v = v_{i} - v_{j}$, is consistently larger than would be
expected from the statistical uncertainties on each individual
velocity measurement ($v_{err} = \sqrt{v_{err,i}^{2} +
  v_{err,j}^{2}}$).  To obtain a standard deviation of $\Delta
v/v_{err}$ of one, we needed to add an additional 0.9~\kms uncertainty
in quadrature with $v_{err}$.  We consider this value of 0.9~\kms to
be the systematic uncertainty in the M2FS velocity measurements, and
we define the total velocity uncertainty for each spectrum to be the
quadrature sum of the Monte Carlo uncertainty from the template fit
and the systematic uncertainty.  A separate check on the systematics
and their possible origin is provided by the high S/N M2FS twilight
sky spectra.  We fit these twilight spectra with a high resolution
solar template spectrum.  The scatter in velocity from fiber to fiber
was $\le 0.20$~\kms, so we conclude that the internal velocity errors
over short timescales on an individual frame (incorporating, e.g., any
fiber-to-fiber systematics) are negligible. However, over multiple
science exposures spanning several hours, this is not necessarily the
case (see above).

In order to verify the reliability of our velocity zero point, we also
measured the velocity of the radial velocity standard star
CD$-43\degr$2527 by fitting it with the HD~122563 template, exactly as
we did for the science spectra.  For the two exposures on
CD$-43\degr$2527 , we find $v_{hel} = 19.6 \pm 0.1$~\kms and $v_{hel}
= 19.9 \pm 0.1$~\kms, compared to the cataloged velocity of $v_{hel} =
19.7 \pm 0.9$~\kms \citep{udry99}.

\subsection{Metallicity Measurements}

We calculated metallicities for 16 \retII RGB stars with the CaT
calibration of \citet{carrera13}.  As recommended by
\citet{hendricks14}, we measured the equivalent widths (EWs) of the
CaT lines in the same way as \citeauthor{carrera13}, fitting each of
the three lines with a Gaussian plus Lorentzian profile.  Also
following \citet{carrera13}, we adopt the line and continuum regions
defined by \citet{cenarro01}, except for the 8498~\AA\ line.
\citeauthor{cenarro01} employed a continuum bandpass of
$8474-8484$~\AA\ for this line, but the blue limit of the GIRAFFE
spectra is 8482~\AA, so we instead use a region on the red side of the
line from $8513-8522$~\AA.  This wavelength range may be modestly
affected by two weak \ion{Fe}{1} lines at 8514~\AA\ and 8515~\AA, but
at the metallicity of typical ultra-faint dwarf stars any resulting
depression of the continuum should be negligible over a 9~\AA\ band.

CaT metallicity measurements usually use the horizontal branch (HB)
magnitude to correct for the dependence of the CaT EWs on stellar
luminosity.  The horizontal branch magnitude of \retII, however, is
not well determined because the galaxy contains so few HB stars.  We
therefore rely on the calibration of CaT EW as a function of absolute
$V$ magnitude from \citet{carrera13}.  We convert the DES $g$ and $r$
magnitudes to the SDSS photometric system, and then use the relations
for metal-poor stars from \citet{jordi06} to transform to $V$.  We
determine absolute magnitudes assuming a distance of $32 \pm 3$~kpc
\citep{bechtol15} and a $V$-band extinction of $A_{V} = 0.05$~mag
\citep{sf11}.

\subsection{Spectroscopic Membership Determination}
\label{membership}

\subsubsection{M2FS}
Out of the 185 M2FS fibers placed on stars, we successfully measured
velocities for 52, including a large majority of the observed targets
brighter than $g=20.6$.  The remaining stars had S/N ratios too low
for spectral features to be confidently detected in the data.  The
velocity measurements and other properties of the stars are listed in
Table~\ref{tab:spec}.  The velocity distribution we measure from the
M2FS spectra exhibits a strong peak at a velocity of $\roughly60$~\kms
(see Fig.~\ref{cmd}), as is characteristic of a gravitationally bound
system.  Approximately half of the stars for which we measure
velocities are contained in this peak, with the remainder spread
across a wide range of heliocentric velocities from $\roughly0$~\kms
to $\roughly330$~\kms.

For a large majority of the observed stars, the membership status is
unambiguous; stars with $v_{hel} > 90$~\kms and
$v_{hel}<40$~\kms are clearly not related to the peak associated with
\retII, while those very near the mean velocity of the system and
close to the central position spatially are almost certainly members.
However, to ensure that the member sample is defined optimally we
carefully examine all stars within 20~\kms of the mean velocity of
\retII, considering their velocities, positions in the color-magnitude
diagram, spatial locations, membership probabilities from
\citet{bechtol15}, and spectral features.  Below we discuss the
individual stars whose membership is not immediately obvious.

\begin{figure}[t]
\epsscale{1.2}
\plotone{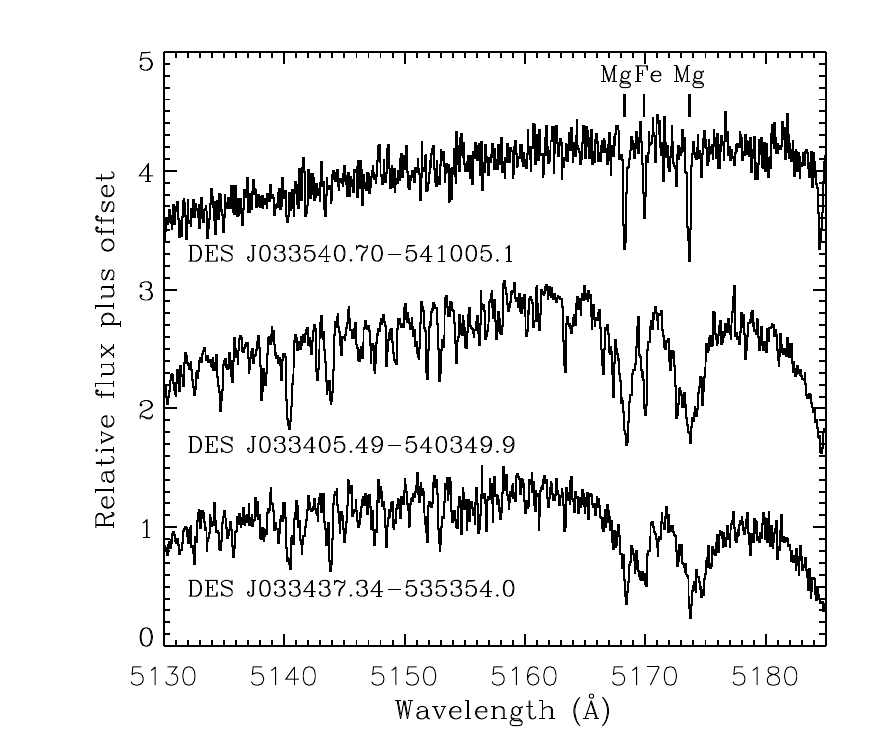}
\caption{Magellan/M2FS spectra in the Mg b triplet region for the
  three stars near the edge of the \retII velocity distribution.  The
  wavelengths of two Mg lines and an Fe line are marked in the bottom
  panel, and the third component of the Mg triplet is just visible at
  a wavelength of 5185~\AA\ at the right edge of each spectrum.  The
  spectrum of DES~J033540.70$-$541005.1 (top) appears similar to that
  of a \retII member, but the color, spatial position, and velocity
  offset of this star make that classification unlikely.  The very
  strong Mg absorption in DES~J033405.49$-$540349.9 (middle), as well
  as the wealth of other absorption features on the blue side of the
  spectrum, indicate that the star is more metal-rich than would be
  expected for a system as small as \retII.  DES~J033437.34$-$535354.0
  (bottom) is a double-lined binary star with a velocity separation of
  $\sim60$~\kms.  The redshifted absorption component from the
  secondary star is most visible in the middle line of the Mg triplet.
}
\label{nonmembers}
\end{figure}

Three stars in our sample have velocities of $v_{hel}\roughly50$~\kms,
just to the left of the \retII peak in Fig.~\ref{cmd}\emph{c}, and
about 15~\kms away from the systemic velocity.  Of these three,
DES~J033405.49$-$540349.9 is very metal-rich,
DES~J033437.34$-$535354.0 is a double-lined spectroscopic binary with
two strong cross-correlation peaks separated by $\sim60$~\kms, and
DES~J033540.70$-$541005.1 is located close to two half-light radii
away from the center of \retII along the minor axis and is separated
from the mean velocity by $\roughly 4\sigma$.  None of these stars has
characteristics consistent with membership in \retII.
DES~J033540.70$-$541005.1 has colors consistent with what would be
expected for a red horizontal branch star in \retII, but is
$\sim0.1$~mag fainter than the isochrone shown in
Fig.~\ref{cmd}\emph{a}.  Given its spatial and velocity offsets from
\retII as well as its position in the color-magnitude diagram, it is
almost certainly not a member.  Spectra of these three non-member
stars are displayed in Fig.~\ref{nonmembers}.

Another group of four stars is located slightly to the blue of the
isochrones we use to describe the \retII stellar population (by $\sim
0.02-0.06$~mag).  DES~J033524.00$-$540226.7 is the farthest from the
isochrone, but is also the faintest of these stars (sitting near the
base of the red giant branch) and consequently has the largest
photometric uncertainties.  Its velocity is within 3~\kms of the
systemic velocity of \retII and it is located inside the
half-light radius, so we consider it a member.
DES~J033635.78$-$540120.2 is also relatively faint, but is farther
away from \retII in both position (at $\roughly 1.3 R_{1/2}$) and
velocity (6~\kms).  It is likely a member of \retII, but the spatial
and velocity offsets make that classification less certain.
DES~J033531.14$-$540148.3 is one of the brightest candidate members,
in the clump of stars near $g \sim 17.5$ in Fig.~\ref{cmd}\emph{a}.
It is within the half-light radius of \retII and has a very low
metallicity as determined from the GIRAFFE CaT measurements.  Because
of its low metallicity and spatial position, we conclude that it is a
\retII member.  The color offset to the blue of the giant branch could
indicate that it is an asymptotic giant branch star.
DES~J033550.10$-$540139.2 is located quite close to the center of
\retII, but is one of the bluest (non-HB) stars with a velocity
consistent with that of \retII.  This star is projected quite close to
the center of \retII and has extremely weak absorption lines.  If it
is indeed a member, it could have an unusually low metallicity.  As a
result of the weak spectral features, the velocity is quite uncertain;
it is consistent with membership, but it has little impact on the
derived properties of \retII because of the large uncertainty.

Two stars are offset to the red side of the isochrone.
DES~J033436.70$-$540645.0 is almost two half-light radii from the
center of \retII, although it is very close to the position of the
velocity peak.  It also has strong absorption lines, indicating a
relatively high metallicity.  Given that it is offset from the rest of
the \retII population in color, spatial position, and metallicity, we
conclude that it is probably not a member.  DES~J033544.18$-$540150.0
is at the centroid of the \retII velocity peak within the
uncertainties, and is also located very close to the central position,
but is $\roughly0.04$~mag redder than the isochrone.  Since it is near
the base of the giant branch, the photometric uncertainties could
contribute to this offset in color, and we consider
DES~J033544.18$-$540150.0 a likely member of \retII.

\begin{figure}[]
\includegraphics[width=\columnwidth]{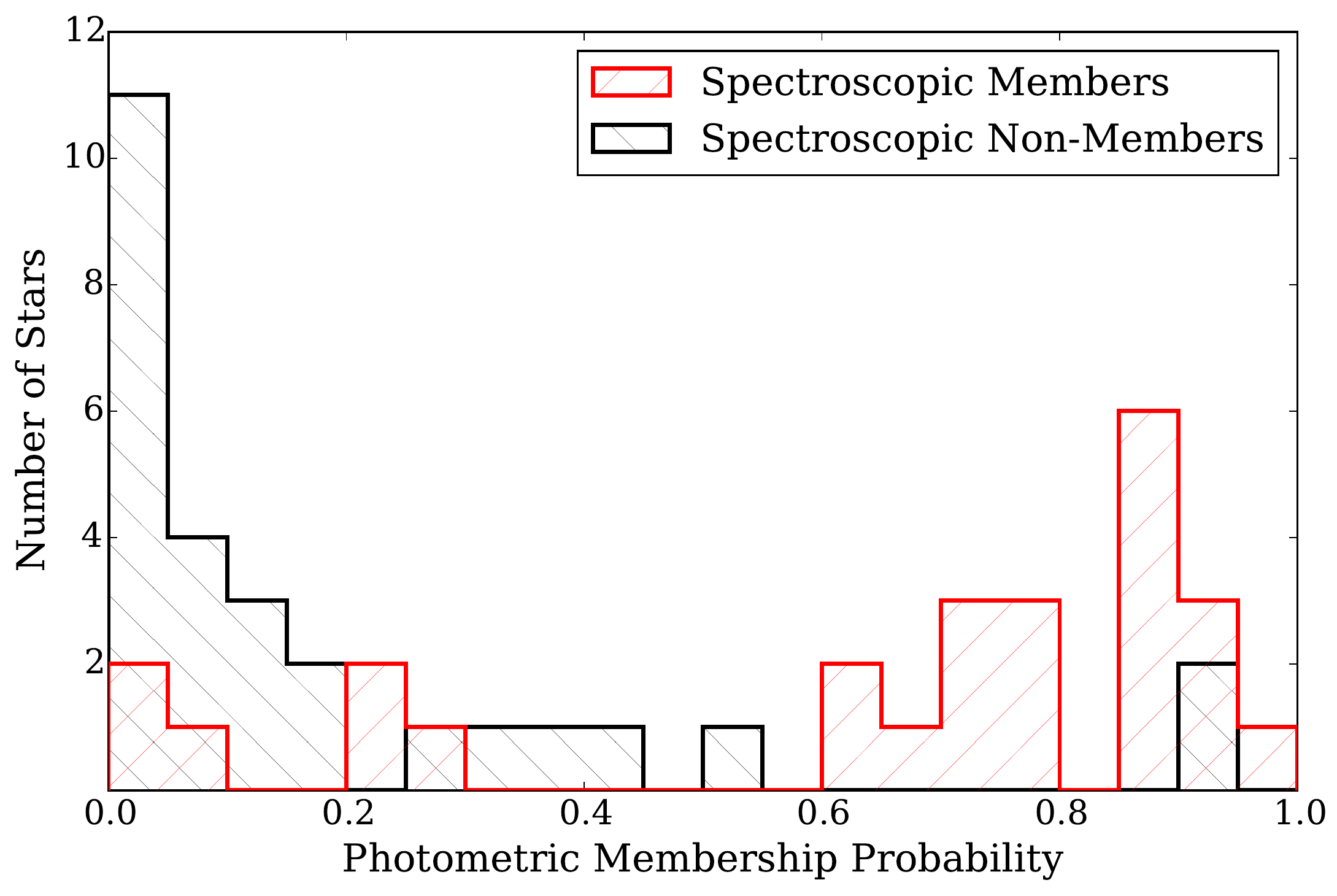}
\caption{Comparison of photometric membership probabilities determined
  from a maximum-likelihood fit to the DES data and spectroscopic
  membership as determined from the velocity measured by M2FS.}
\label{fig:mem}
\end{figure}

Because the stars for which membership is plausible have velocities
quite similar to that of \retII (and in some cases have large
uncertainties), including or excluding them from the member sample
does not have any significant effect on the properties we derive for
\retII in \S\ref{discussion}.  We show the correspondence between M2FS
spectroscopic members and high photometric membership probability in
Fig.~\ref{fig:mem}.

\subsubsection{GIRAFFE and GMOS}
\label{giraffe_gmos_mem}
We also identify a handful of \retII members in the GIRAFFE and GMOS
data sets that were not observed with M2FS.  We use a velocity
measurement based on the Paschen lines to confirm that the candidate
BHB star DES~J033539.85$-$540458.1 (\S\ref{hb}) observed by GMOS is
indeed a member of \retII, with a velocity of $69 \pm 6$~\kms.  The
GIRAFFE targets included a bright ($g \sim 16.5$) star at
$(\alpha_{2000}, \delta_{2000}) = ({\rm 03{:}35{:}23.85,
  -54{:}04{:}07.5})$ that was omitted from our photometric catalog and
M2FS observations because it is saturated in the coadded DES images.
However, the spectrum of the star makes clear that it is very
metal-poor and is within a few \kms of the systemic velocity of
\retII.  While the magnitudes derived from individual DES frames place
it slightly redder than the isochrone that best matches the lower red
giant branch of \retII, it is also located inside the half-light
radius, and is very likely a member.  In fact, it is probably the
brightest star in any of the ultra-faint dwarfs.

Another GIRAFFE target, DES\,J033548.04$-$540349.8, was not included
in the M2FS sample because it was assigned a photometric membership
probability of zero by the maximum-likelihood analysis
\citep{bechtol15}. It is very slightly redder (relative to the
isochrone) than the bulk of the \retII stars, but it is located within
the half-light radius and has a velocity $\roughly1$~\kms from the
systemic velocity.  The metallicity we determine from the GIRAFFE
spectrum is low, but somewhat more metal-rich than any of the other
member stars, perhaps consistent with its color.  The
\ion{Mg}{1}~$\lambda8807$ line advocated by \citet{bs12} as a
discriminant between dwarf-galaxy red giants and foreground main
sequence stars is weak (${\rm EW} = 0.1$~\AA), suggesting that it is a
member of \retII, but since it would be the highest metallicity star
in the galaxy we regard this conclusion as tentative.  Although all
three of these stars are likely members of \retII, we do not include
them in the determination of the kinematic properties in
\S\ref{sigma} because they have not been placed accurately enough in
the same velocity reference frame.

We classify the GIRAFFE target DES\,J033524.96$-$540230.7 as a non-member
based on its extremely broad CaT lines, despite its overlap with the
\retII velocity distribution.

Finally, we note that for DES\,J033515.17$-$540843.0, the M2FS and
GIRAFFE velocity measurements differ by $8.2 \pm 3.5$~\kms.  The
significance of this difference is $2.4 \sigma$, corresponding to a
p-value of 2\%.  While we cannot rule out the possibility that this
difference in velocity is attributable to chance or systematics, we
suggest that this star may be a binary system in \retII.  Orbital
motion of 8~\kms over a time span of 18 days is consistent with the
handful of previously detected binaries in ultra-faint dwarf galaxies
\citep{koposov11,koch14}.

\subsubsection{Foreground Modeling}
\label{foreground}
Finally, we consider the expected contamination from foreground stars
in a statistical sense.  We use the Besan{\c c}on \citep{besancon} and
TRILEGAL \citep{girardi05,vanhollebeke09} Galactic stellar models to
calculate the expected velocity distribution of foreground stars in
the region of \retII.  We select stars within $0.02 \magn$ of the
composite isochrone (\S\ref{sec:targets}) and with $17 < g < 20.6$.
From the resulting velocity distribution, we find that $\roughly 10\%$
of foreground stars passing these photometric cuts have a velocity
consistent with the heliocentric velocity peak of \retII (55 --
70\,\kms).  We set the overall normalization for the number of
foreground stars by assuming that all stars in the sideband of the
peak ($v < 40 \kms$ or $v > 90\kms$) belong to the foreground
distribution.  The expected number of foreground stars within the peak
of ranges from 3.0 (Besan{\c c}on) to 4.1 (TRILEGAL).  Given the small
number statistics, this is consistent with the results of the
membership analysis above.  Under these assumptions, the
foreground-only hypothesis for the \retII velocity peak is excluded
at $p \approx 10^{-12}$.

\subsection{Horizontal Branch Stars}
\label{hb}

The DES photometric catalog contains three candidate blue HB (BHB)
stars, three candidate red HB stars, and one star located slightly
below the HB at an intermediate color that could be consistent with
the RR~Lyrae instability strip.  We targeted all seven of these stars
with M2FS, but were only able to determine velocities for five of
them.  For the BHB stars, we confirm DES~J033618.68$-$535745.1 as a
member of \retII, but we do not detect any spectral features for
DES~J033539.85$-$540458.1, and DES~J033612.7$-$535602.2 was observed
with a fiber with 3\%\ throughput so the spectrum was not useful.  As
mentioned in \S\ref{membership}, our GMOS spectrum of
DES~J033539.85$-$540458.1 confirms it as a \retII member.  The
velocities of the three RHB candidates indicate that they are not
members.  The final HB candidate, DES~J033341.71$-$540007.3, has a
measured velocity of $34.4 \pm 1.2$~\kms, which could be consistent
with membership in \retII if it were an RR~Lyrae variable, but we do
not detect photometric variations for it or any of the BHB stars in
the $2-4$ DES single-epoch images per filter available in this field.

\begin{figure*}[th!]
\epsscale{1.2}
\plotone{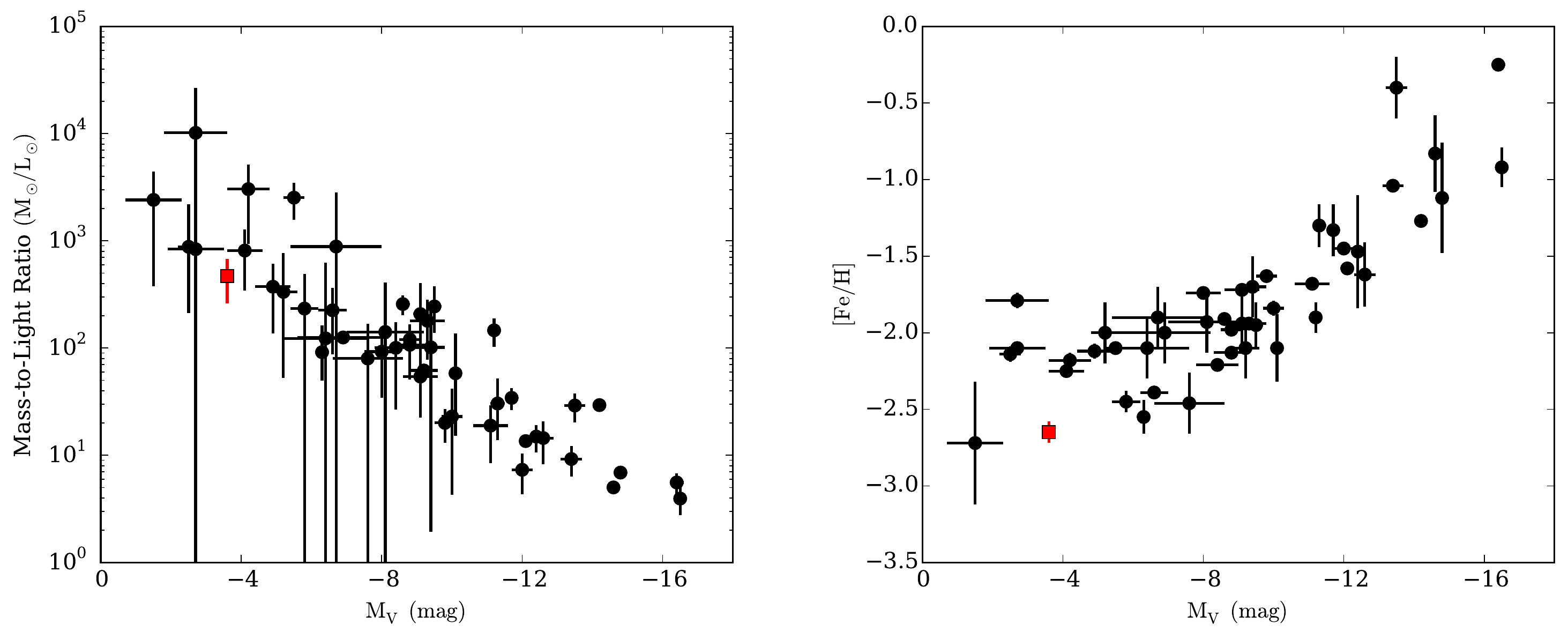}
\caption{Mass-to-light ratio \emph{(left)} and metallicity
  \emph{(right)} of \retII\ compared to other galaxies within 1\Mpc.
  Mass-to-light ratios were calculated from the velocity dispersions,
  half-light radii, and absolute magnitudes compiled by
  \citet{McConnachie:2012vd} in order to maintain consistency with the
  value we report for \retII.  Metallicities were also adopted from
  the \citet{McConnachie:2012vd} compilation, with updates from
  \citet{kirby13} where newer measurements were available.}
\label{m2l_feh}
\end{figure*}

\section{DISCUSSION}
\label{discussion}

\subsection{Velocity Dispersion and Mass}
\label{sigma}

After identifying likely members of \retII with a combination of
photometry, velocity, and metallicity information, we can calculate
the systemic velocity and velocity dispersion of \retII.  For the
sample of 25 member stars with M2FS velocity measurements identified
in \S\ref{measurements}, we find a systemic velocity of $v_{hel} =
\vbulk$.  Using the method of \citet{walker06}, we determine a
maximum-likelihood estimate for the velocity dispersion of $\sigma_v =
\vdisp$.  If we instead rely on the GIRAFFE data set (which includes
17 member stars), we obtain fully compatible values of $v_{hel} =
\ensuremath{63.3 \pm 0.8 \kmss}\xspace$ and $\sigma_v =
\ensuremath{3.5 \pm 1.0 \kmss}\xspace$.  The velocity dispersion of
\retII is also robust to membership uncertainties; decisions about
including or excluding the plausible member stars discussed in
\S\ref{membership} change the calculated dispersion by less than the
stated uncertainty.  Only by adding stars that are very unlikely to be
members for multiple reasons can the best-fit velocity dispersion be
significantly increased.  In principle, the low velocity dispersion of
\retII increases the likelihood that the measured dispersion could be
artificially inflated because of the orbital motions of binary stars.
While our observations do not span a long enough time baseline to have
much sensitivity to binaries, previous studies have shown that even in
the smallest dwarfs binary stars do not significantly inflate the
observed velocity dispersion \citep{simon11,martinez11}.

\citet{wolf10} showed that, independent of the velocity anisotropy of
a stellar system, the mass enclosed within the half-light radius can
be accurately computed as
\begin{equation}
M_{1/2} = 930 \left( \frac{\sigma_v^{2}}{{\rm km^{2}~s^{-2}}} \right)
\left( \frac{\rhalf}{{\rm pc}} \right) {\rm M}_{\odot}.
\end{equation}

\noindent
Using this relation, the mass within the elliptical half-light radius
of \retII \citep[$\rhalf = 55 \pm 5\pc$;][]{bechtol15} is \mass.  The
absolute magnitude of $M_{V} = -3.6 \pm 0.1$ from \citet{bechtol15}
translates to a luminosity of $2360 \Lsun$, leading to a mass-to-light
ratio of \masstolight.  This value is consistent with the inverse
correlation between mass-to-light ratio and luminosity for other Local
Group dwarf galaxies, although \retII is on the low end for galaxies
of similar luminosity (see Fig.~\ref{m2l_feh}).

\subsection{Metallicity Spread}

\begin{figure*}[t]
\epsscale{1.2}
\plotone{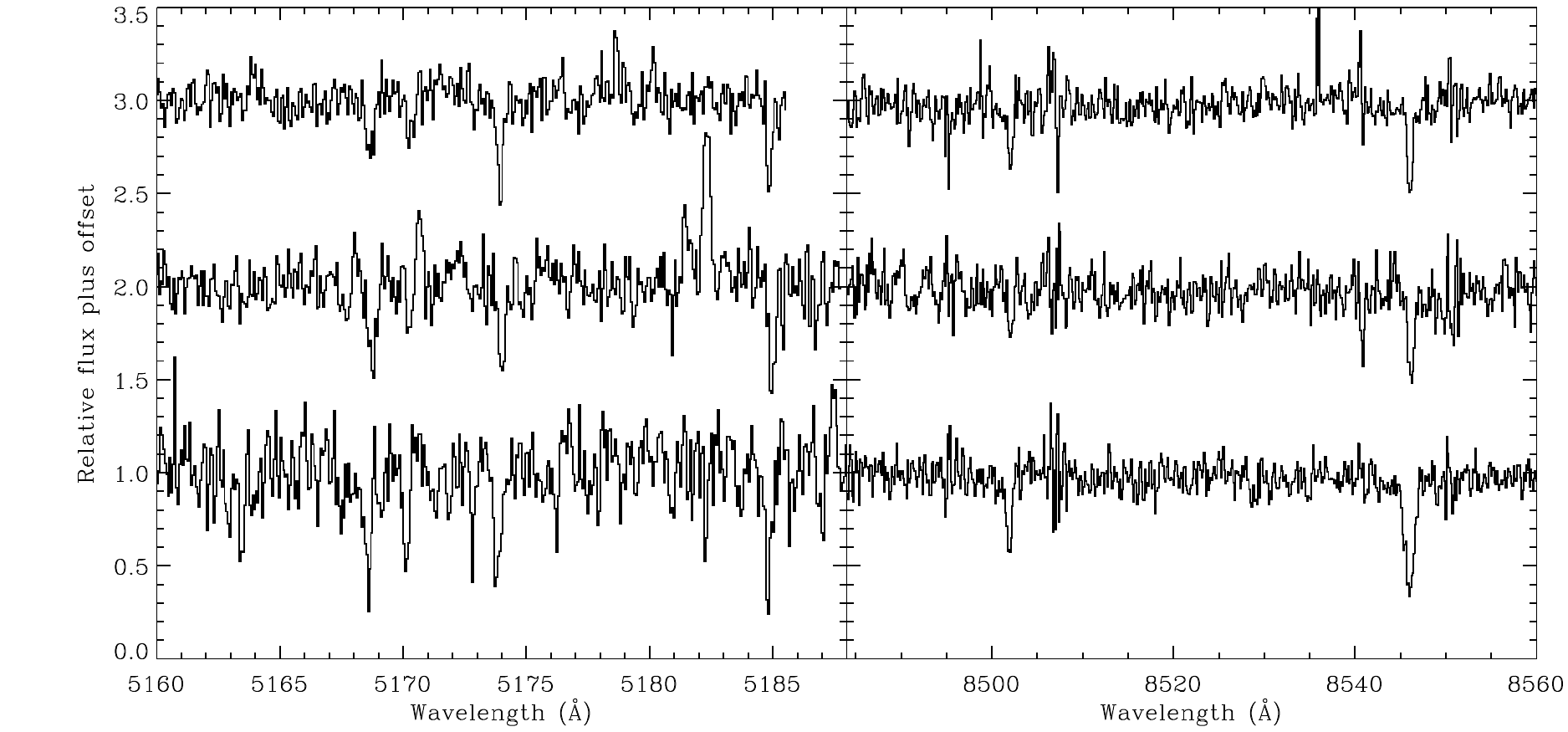}
\caption{\emph{(Left)} Magellan/M2FS spectra in the Mg b triplet
  region for three \retII member stars covering a range of line
  strengths.  From top to bottom, the stars are
  DES~J033556.28$-$540316.3, DES~J033454.24$-$540558.0, and
  DES~J033457.57$-$540531.4.  These stars span only 0.1~mag in
  luminosity and 0.08~mag in $g-r$ color, so their effective
  temperatures and surface gravities should be very similar.  Any
  differences in line strength therefore translate directly into
  chemical abundance differences.  The apparent emission features near
  5182~\AA\ in the spectrum of DES~J033454.24$-$540558.0 are
  contamination by the Littrow ghost \citep{burgh07}.  \emph{(Right)}
  VLT/GIRAFFE spectra of the bluer two CaT lines for the same
  stars.}
\label{members}
\end{figure*}

Even a visual inspection of the M2FS spectra reveals clear chemical
abundance differences within the \retII member sample.  In
Fig.~\ref{members} we display spectra of three of the stars within the
clump of \retII members at $g \approx 18.9$ in Fig.~\ref{cmd}\emph{a}.
The absolute strength of the Mg absorption and the Fe absorption, as
well as the ratio between the two, varies from star to star,
demonstrating that \retII is not chemically homogeneous.

\begin{figure*}[H!]
\epsscale{1.2}
\plotone{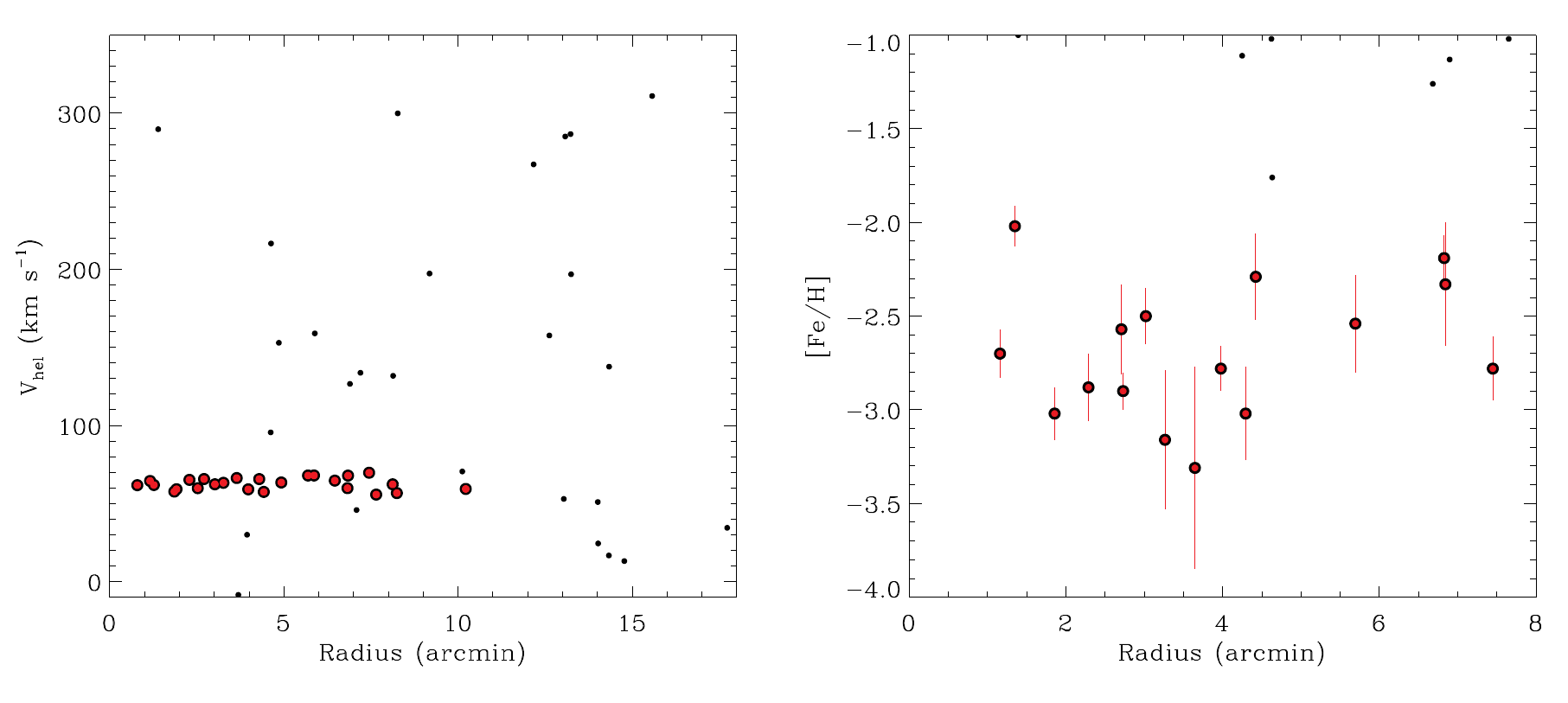}
\caption{\emph{(Left)} Velocity as a function of radius for the M2FS
  spectroscopic sample.  \retII members are plotted as filled red
  symbols, and non-members are shown as small black dots.
  \emph{(Right)} Metallicity as a function of radius for the GIRAFFE
  spectroscopic sample; symbols are the same as in the left panel.
  Note that CaT metallicity measurements depend on the absolute
  magnitude of the star, so for non-member stars the displayed points
  do not represent the true metallicity because those stars are not at
  the distance of Ret~II.}
\label{r_v_feh}
\end{figure*}

The CaT metallicity measurements from the VLT/GIRAFFE spectra provide
a more quantitative demonstration of the spread in overall metallicity
from star to star in \retII.  We find member stars spanning more than
1~dex in iron abundance, ranging from $\rm{[Fe/H]} = -2.0$ to
$\rm{[Fe/H]} = -3.3$.  The individual metallicities are displayed as a
function of radius in Fig.~\ref{r_v_feh}\emph{b}.  There is no
indication of a metallicity gradient within \retII.  The metallicity
spread, calculated with the same technique we used in \S\ref{sigma}
for the velocity dispersion, is $0.28 \pm 0.09$~dex.  If we instead
employ the definition of \citet{kirby11}, we find a similar spread.
Both this dispersion and the presence of several extremely metal-poor
stars at $\rm{[Fe/H]} < -3$ demonstrate that \retII has the chemical
characteristics of a dwarf galaxy rather than a globular cluster.  The
mean metallicity of \retII is $\rm{[Fe/H]} = \feh$, ranking it along
with Segue~1 as the most metal-poor galaxies known.  This metallicity
measurement places \retII somewhat below the stellar mass-metallicity
relationship determined by \citet{kirby13} and argues against the
existence of a metallicity floor for dwarf galaxies
(Fig.~\ref{m2l_feh}).  The kinematic and chemical properties of
  \retII are summarized in Table~\ref{ret2_table}.

\begin{deluxetable}{llr}
%\tabletypesize{\scriptsize}
\tablecaption{Summary of Properties of Reticulum\,II}
\tablewidth{0pt}
\tablehead{
\colhead{Row} & \colhead{Quantity} & \colhead{Value}
}
\startdata
(1) & RA (J2000)                           & 03:35:41 \\
(2) & Dec (J2000)                          & $-54$:03:00 \\
(3) & Distance (kpc)                       & $32$  \\
(4) & $M_{V,0}$                             & $-3.6 \pm 0.1$ \\
(5) & $L_{V,0}$ (L$_{\odot}$)               & $2360 \pm 200$ \\
(6) & $\epsilon$                           & $0.60^{+0.10}_{-0.20}$ \\
(7) & $r_{\rm 1/2}$ (pc)                    & $55 \pm 5$  \\
\hline
(8)  & $V_{hel}$ (\kms)                     & $62.8 \pm 0.5$ \\
(9)  & $V_{\rm GSR}$ (\kms)                 & $-92.5 \pm 0.5$ \\
(10)  & $\sigma$ (\kms)                    & $3.3 \pm 0.7$ \\
(11)  & Mass (M$_{\odot}$)                  & $5.6 \pm 2.4 \times 10^{5}$  \\
(12)  & M/L$_{V}$ (M$_{\odot}$/L$_{\odot}$)  & $470 \pm 210$ \\
(13)  & Mean [Fe/H]                        & $-2.65 \pm 0.07$ \\
(14)  & Metallicity dispersion (dex)       & $0.28 \pm 0.09$ \\
(15)  & $\log_{10}{J(0.2\degr)}$ (GeV$^{2}$~cm$^{-5}$) & $18.8 \pm 0.6$ \\
(16)  & $\log_{10}{J(0.5\degr)}$ (GeV$^{2}$~cm$^{-5}$) & $18.9 \pm 0.6$
\enddata

\tablecomments{Rows (1)-(7) are taken from the DES photometric
  analysis of \citet{bechtol15}.  Values in rows (8)-(16) are derived
  in this paper.}
\label{ret2_table}
\end{deluxetable}

\subsection{Reticulum II and the Magellanic Clouds}

The proximity of the DES satellites to the Magellanic Clouds suggests
that some or all of them may have originated in a Magellanic group.
\retII is the closest of these objects to the Magellanic Clouds, with
a three-dimensional separation of just $23 \pm 3$~kpc.  The radial
velocity of \retII differs from that of the LMC by 200~\kms, which
represents a lower limit on the difference in their space motions.
The total mass of the LMC out to a distance of 23~kpc is not known,
but \citet{kallivayalil13} argue that reasonable bounds are $3 \times
10^{10} \Msun$ to $2.5 \times 10^{11} \Msun$.  Using DES imaging,
\citet{balbinot15} determine a truncation radius of $13.5 \pm 0.08
\kpc$ for the LMC.  If this truncation is tidal in origin, the favored
mass is at the lower limit of the range suggested by
\citet{kallivayalil13}.  For a mass of $3 \times 10^{10} \Msun$, the
escape velocity of the LMC at $d = 23$~kpc is only 106~\kms, and for
$M_{\rm{LMC}} < 10^{11}\Msun$ the escape velocity is less than
$200\kms$.  Only if the LMC mass is higher than this value and the
tangential velocity of \retII is similar to that of the LMC could
\retII currently be a bound satellite of the Magellanic Clouds.  Of
course, even if \retII is not bound to the LMC now, it is possible
that it was an LMC satellite at earlier times and was stripped from
that host by the tidal field of the Milky Way more recently.  A proper
motion measurement of \retII will be needed to conclusively determine
its origin, and its location just 32~kpc away from the Sun should
enable the proper motion to be determined in the next few years.

The observed velocity distribution of high velocity ($v_{hel} >
100$~\kms) stars displayed in Figure~\ref{cmd}\emph{c} appears
inconsistent with the distribution of velocities for Milky Way stars
predicted by Galactic models (\S\ref{foreground}).  In particular,
there are unexpected concentrations of stars around 145~\kms (six
stars) and 300~\kms (five stars).  The former group of stars is near
the heliocentric velocity of the SMC \citep[145.6\kms;][]{hz06},
suggesting that these might be distant SMC stars.  If they are at the
distance of the SMC, they are at projected separations of 27~kpc,
approximately twice as far out as the most distant currently known SMC
population \citep{nidever11}, and have therefore likely been tidally
stripped.  The higher velocity stars have very similar velocities to
the Magellanic Stream gas a few degrees away from \retII, and could
represent a stellar counterpart of the Stream.

\subsection{J-Factor}

It is posited that dark matter particles could self-annihilate to
produce gamma rays \citep[e.g.,][]{gunn78,bergstrom88,Baltz:2008wd}.
The large dark matter content, relative proximity, and low
astrophysical foregrounds of dwarf galaxies make them promising
targets for the detection of these gamma rays.  The predicted signal
from the annihilation of dark matter particles is proportional to the
line-of-sight integral through the square of the dark matter density
\citep[\eg,][]{Baltz:2008wd},
\begin{equation}
\label{eq:jfactor}
J(\Delta \Omega) = \int_{\Delta\Omega}\int_{\rm l.o.s.}\rho_{\rm DM}^{2}(\vect{r})\,\text{d}s\,\text{d}\Omega'.
\end{equation}  
Here, $\rho_{\rm DM}(\vect{r})$ is the dark matter particle density,
and the integral is performed over a solid angle $\Delta \Omega$.  The
\Jfactor is derived by modeling the velocities using the spherical
Jeans equation, with assumptions on the theoretical priors for the
parameters that describe the dark matter halo
\citep[\eg,][]{Strigari:2007at,Essig:2009jx,Charbonnier:2011ft,Martinez:2013ioa,Geringer-Sameth:2014yza}.
Here, we model the dark matter halo as a generalized
Navarro-Frenk-White (NFW) profile \citep{Navarro:1996gj}.  We use
flat, `uninformative' priors on the dark matter halo parameters
\citep[see][]{Essig:2009jx} and assume a constant stellar velocity
anisotropy.  Using this procedure, we find an integrated \Jfactor for
\retII of $\log_{10}(J) = \jsmall$ within an angular cone of radius
$0.2\degr$, and $\log_{10}(J) = \jlarge$ within $0.5\degr$.  This
latter value assumes that the dark matter halo extends beyond the
radius of the outermost spectroscopically confirmed star, but
truncates within the estimated tidal radius for the dark matter halo
($\roughly 1$~kpc). The quoted uncertainties are $1\sigma$, and are
estimated by modeling the posterior probability density function of
$\log_{10}(J)$ as a Gaussian.  Note that the uncertainty obtained by
modeling this individual system is larger than is obtained by modeling
the entire population of dSphs \citep{Martinez:2013ioa}.

Several previously known ultra-faint dwarf galaxies possess larger
mean \Jfactors than \retII, most notably Segue~1, Ursa~Major~II, and
Coma~Berenices
\citep{Ackermann:2013yva,Geringer-Sameth:2014yza,Conrad:2015bsa}.
Though the velocity dispersions of \retII and Segue~1 are consistent
within uncertainties, \retII is more distant ($32\kpc$ compared to
$23\kpc$) and has a larger half-light radius as measured along the
major axis ($55\pc$ compared to $29\pc$). The larger distance and
larger half-light radius imply a reduced mean \Jfactor relative to
Segue~1. In comparison to Ursa~Major~II, \retII is at a similar
distance, but has a velocity dispersion that is smaller by roughly a
factor of two.  The larger dispersion, and hence mass, accounts for
the larger \Jfactor of Ursa~Major~II. Coma~Berenices is more distant
than \retII ($44 \kpc$ compared to $32 \kpc$); however, the larger
velocity dispersion of Coma~Berenices implies a slightly larger mean
\Jfactor.

Since Segue~1, Ursa~Major~II, and Coma~Berenices all possess larger
\Jfactors than \retII, we expect dark matter annihilation to produce a
larger gamma-ray flux from these objects.  However, no gamma-ray
excess has been associated with any of the previously known dwarf
galaxies \citep{Ackermann:2015zua}.  Given comparable gamma-ray
sensitivity, it is unlikely that a dark matter annihilation signal
would be detected from \retII without also being detected from dwarf
galaxies with higher \Jfactors
\citep{dw15,Geringer-Sameth:2015lua,Hooper:2015ula}.

\section{SUMMARY AND CONCLUSIONS}
\label{conclusions}

We have presented the first spectroscopic analysis of the recently
discovered Milky Way satellite Reticulum~II.  We measure the
velocities of 25 \retII members using high resolution spectroscopy
from Magellan/M2FS, as well as metallicities from the Ca triplet lines
for 16 members using high resolution spectroscopy from VLT/GIRAFFE and
6 members using low resolution spectroscopy from Gemini South/GMOS.
\retII has a velocity dispersion of $\sigma_v = \vdisp$, corresponding
to a dynamical mass within its half-light radius of \mass and a
mass-to-light ratio of \masstolight.

The metallicity of \retII determined from the CaT is ${\rm [Fe/H]} =
\feh$, consistent with that of Segue~1 \citep{fsk14} within the
uncertainties, and $\sim0.2$~dex lower than that of any other known
galaxy.  We find that \retII has an internal metallicity spread of
$0.28 \pm 0.09$~dex, with stars spanning a total range of more than
1~dex.  Even the most metal-rich stars in the galaxy are at ${\rm
  [Fe/H]} \approx -2.0$, and \retII contains 4 extremely metal-poor
stars with ${\rm [Fe/H]} < -3$.  Both its chemical and kinematic
properties confirm that \retII is a dwarf galaxy.

The location of \retII just 23~kpc away from the LMC suggests that it
could have originated as a satellite of the Magellanic system rather
than having always been associated with the Milky Way
\citep{bechtol15,koposov15}.  However, our measured systemic radial
velocity of $v_{hel} = \vbulk$ means that \retII is moving away from
the LMC at a minimum velocity of 199\kms.  According to current LMC
mass estimates, this velocity likely exceeds the escape velocity of
the LMC, indicating that the two objects are not gravitationally
bound.  This result does not rule out the possibility that \retII was
previously a Magellanic satellite, and future proper motion
measurements will shed more light on its origin.

The \Jfactor calculated from the internal kinematics of \retII is
$\log_{10}{J} = \jsmall$ within a radius of $0.2\degr$, somewhat lower
than previously estimated based on the galaxy's distance alone
\citep{dw15}.  The predicted gamma-ray flux from dark matter
annihilation in \retII is therefore likely to be lower than that
predicted for several other Milky Way satellites
\citep{dw15,Geringer-Sameth:2015lua,Hooper:2015ula}.

Satellite galaxies like \retII provide a crucial testing ground for
the $\Lambda$CDM paradigm, and accordingly, the search for ultra-faint
galaxies has become a major theme of near-field cosmology.  It is
expected that many additional Milky Way satellite galaxies could be
found in ongoing and near-future wide-field optical imaging surveys
\citep{Tollerud:2008ze,Hargis:2014kaa}.  The link between newly
discovered stellar systems and the dark matter halos in which they may
reside is established by follow-up dynamical and chemical analysis.
\retII is the first of several recently reported stellar systems
\citep{bechtol15,koposov15,laevens15,martin15,kim15} to be
spectroscopically confirmed as a dark-matter-dominated Milky Way
satellite galaxy.  The spectroscopic campaign to characterize new
satellite galaxy candidates represents an essential step in ongoing
tests of the standard cosmological model.

\tabletypesize{\scriptsize}
\begin{deluxetable*}{ccccccccccccc}[th!]
\tablecaption{Velocity and metallicity measurements.\label{tab:spec}}

\tablehead{ID & RA & DEC & $g$\tablenotemark{a} & $r$\tablenotemark{a} & $v$ & ${\rm EW_\textsc{GIRAFFE}}$ & ${\rm [Fe/H]_\textsc{GIRAFFE}}$ & ${\rm EW_\textsc{GMOS}}$ & ${\rm [Fe/H]_\textsc{GMOS}}$ & MEM \\ 
 & (deg) & (deg) & (mag) & (mag) & (\kms) &  &  &  &  &  }
\startdata

\\ \multicolumn{11}{c}{ \ruleline{M2FS} } \\
DES\,J033341.71$-$540007.3 & 53.42377 & -54.00201 & 18.26 & 18.20 & $34.42 \pm 1.16$ & -- & -- & -- & -- &   NM\\
DES\,J033405.49$-$540349.9 & 53.52287 & -54.06387 & 17.58 & 16.99 & $50.96 \pm 0.94$ & -- & -- & -- & -- &   NM\\
DES\,J033413.06$-$535956.1 & 53.55441 & -53.99892 & 19.45 & 18.99 & $196.81 \pm 1.44$ & -- & -- & -- & -- &   NM\\
DES\,J033413.94$-$540934.4 & 53.55807 & -54.15956 & 18.26 & 17.68 & $16.70 \pm 1.06$ & -- & -- & -- & -- &   NM\\
DES\,J033418.32$-$541006.2 & 53.57632 & -54.16839 & 18.02 & 17.45 & $24.41 \pm 0.98$ & -- & -- & -- & -- &   NM\\
DES\,J033429.94$-$541111.8 & 53.62474 & -54.18661 & 19.77 & 19.32 & $286.61 \pm 1.55$ & -- & -- & -- & -- &   NM\\
DES\,J033430.07$-$540922.2 & 53.62528 & -54.15618 & 18.17 & 17.85 & $267.19 \pm 0.97$ & -- & -- & -- & -- &   NM\\
DES\,J033436.70$-$540645.0 & 53.65291 & -54.11249 & 18.85 & 18.29 & $70.52 \pm 1.01$ & -- & -- & -- & -- &  NM?\\
DES\,J033437.34$-$535354.0 & 53.65557 & -53.89832 & 17.69 & 17.11 & $52.94 \pm 1.10$ & -- & -- & -- & -- &   NM\\
DES\,J033437.98$-$541359.7 & 53.65825 & -54.23324 & 20.55 & 20.07 & $137.64 \pm 2.58$ & -- & -- & -- & -- &   NM\\
DES\,J033439.66$-$540754.4 & 53.66523 & -54.13177 & 19.38 & 18.87 & $59.32 \pm 1.84$ & -- & -- & -- & -- &    M\\
DES\,J033439.81$-$540058.5 & 53.66587 & -54.01626 & 18.21 & 17.93 & $197.28 \pm 1.00$ & -- & -- & -- & -- &   NM\\
DES\,J033447.94$-$540525.0 & 53.69974 & -54.09028 & 17.52 & 16.92 & $62.28 \pm 1.03$ & -- & -- & -- & -- &    M\\
DES\,J033449.20$-$535019.7 & 53.70502 & -53.83881 & 17.62 & 17.02 & $13.13 \pm 0.93$ & -- & -- & -- & -- &   NM\\
DES\,J033453.23$-$541403.9 & 53.72180 & -54.23443 & 20.59 & 20.19 & $285.07 \pm 2.73$ & -- & -- & -- & -- &   NM\\
DES\,J033453.50$-$540454.6 & 53.72291 & -54.08182 & 20.18 & 19.80 & $133.74 \pm 2.58$ & -- & -- & -- & -- &   NM\\
DES\,J033454.24$-$540558.0 & 53.72600 & -54.09945 & 18.95 & 18.44 & $69.70 \pm 1.38$ & $1.69 \pm 0.25$ & $-2.78 \pm 0.17$ & -- & -- &    M\\
DES\,J033457.57$-$540531.4 & 53.73988 & -54.09206 & 18.94 & 18.42 & $59.84 \pm 1.17$ & $2.77 \pm 0.23$ & $-2.19 \pm 0.12$ & -- & -- &    M\\
DES\,J033502.50$-$540354.3 & 53.76041 & -54.06507 & 19.24 & 18.75 & $67.94 \pm 1.14$ & $1.98 \pm 0.43$ & $-2.54 \pm 0.26$ & -- & -- &    M\\
DES\,J033502.87$-$540109.8 & 53.76196 & -54.01940 & 20.36 & 19.93 & $67.96 \pm 3.48$ & -- & -- & -- & -- &    M\\
DES\,J033506.56$-$540604.3 & 53.77734 & -54.10120 & 19.79 & 19.28 & $158.97 \pm 1.36$ & -- & -- & -- & -- &   NM\\
DES\,J033509.50$-$540229.7 & 53.78959 & -54.04158 & 18.24 & 17.76 & $95.55 \pm 0.96$ & $5.72 \pm 0.29$ & -- & -- & -- &   NM\\
DES\,J033511.66$-$540321.8 & 53.79858 & -54.05606 & 19.31 & 18.83 & $65.64 \pm 1.34$ & $1.27 \pm 0.27$ & $-3.02 \pm 0.25$ & -- & -- &    M\\
DES\,J033513.73$-$540456.7 & 53.80722 & -54.08242 & 19.70 & 19.24 & $57.43 \pm 2.38$ & $2.26 \pm 0.40$ & $-2.29 \pm 0.23$ & -- & -- &    M\\
DES\,J033514.01$-$540558.2 & 53.80839 & -54.09949 & 20.01 & 19.57 & $63.47 \pm 1.45$ & -- & -- & -- & -- &    M\\
DES\,J033515.17$-$540843.0 & 53.81322 & -54.14529 & 19.73 & 19.25 & $67.87 \pm 1.36$ & $2.19 \pm 0.58$ & $-2.33 \pm 0.33$ & -- & -- &    M\\
DES\,J033517.01$-$540403.0 & 53.82088 & -54.06751 & 19.72 & 19.24 & $66.31 \pm 1.40$ & $0.93 \pm 0.38$ & $-3.31 \pm 0.54$ & $1.37 \pm 0.20$ & $-2.86 \pm 0.18$ &    M\\
DES\,J033520.37$-$541816.9 & 53.83486 & -54.30470 & 20.53 & 20.18 & $311.02 \pm 2.83$ & -- & -- & -- & -- &   NM\\
DES\,J033520.97$-$540348.2 & 53.83736 & -54.06338 & 18.95 & 18.46 & $62.31 \pm 1.06$ & $2.16 \pm 0.26$ & $-2.50 \pm 0.15$ & $1.15 \pm 0.28$ & $-3.20 \pm 0.30$ &    M\\
DES\,J033524.00$-$540226.7 & 53.85002 & -54.04075 & 20.31 & 19.89 & $59.81 \pm 1.83$ & -- & -- & -- & -- &    M\\
DES\,J033531.14$-$540148.2 & 53.87975 & -54.03007 & 17.64 & 17.07 & $57.66 \pm 0.96$ & $1.72 \pm 0.20$ & $-3.02 \pm 0.14$ & $1.88 \pm 0.26$ & $-2.92 \pm 0.16$ &    M\\
DES\,J033533.71$-$535025.4 & 53.89045 & -53.84038 & 18.75 & 18.25 & $157.61 \pm 2.22$ & -- & -- & -- & -- &   NM\\
DES\,J033535.44$-$540254.9 & 53.89766 & -54.04857 & 20.68 & 20.35 & $61.74 \pm 4.79$ & -- & -- & -- & -- &    M\\
DES\,J033536.94$-$535445.1 & 53.90391 & -53.91253 & 20.56 & 20.16 & $299.87 \pm 2.00$ & -- & -- & -- & -- &   NM\\
DES\,J033537.06$-$540401.2 & 53.90442 & -54.06701 & 18.57 & 18.03 & $64.44 \pm 1.14$ & $1.94 \pm 0.21$ & $-2.70 \pm 0.13$ & $1.67 \pm 0.48$ & $-2.86 \pm 0.32$ &    M\\
DES\,J033540.70$-$541005.1 & 53.91957 & -54.16809 & 18.23 & 17.92 & $45.83 \pm 0.99$ & -- & -- & -- & -- &   NM\\
DES\,J033544.18$-$540150.0 & 53.93409 & -54.03056 & 20.40 & 19.91 & $61.86 \pm 1.99$ & -- & -- & -- & -- &   M?\\
DES\,J033546.17$-$540733.9 & 53.94235 & -54.12608 & 19.23 & 18.74 & $216.54 \pm 1.16$ & $3.51 \pm 0.41$ & -- & -- & -- &   NM\\
DES\,J033547.47$-$535926.5 & 53.94781 & -53.99068 & 21.25 & 20.99 & $-8.53 \pm 4.78$ & -- & -- & -- & -- &   NM\\
DES\,J033549.96$-$540321.5 & 53.95816 & -54.05596 & 19.66 & 19.21 & $289.73 \pm 1.23$ & $4.93 \pm 0.36$ & -- & $4.65 \pm 0.59$ & -- &   NM\\
DES\,J033550.10$-$540139.2 & 53.95873 & -54.02756 & 19.69 & 19.26 & $59.15 \pm 8.25$ & -- & -- & -- & -- &   M?\\
DES\,J033552.08$-$540733.9 & 53.96700 & -54.12608 & 20.56 & 20.21 & $152.93 \pm 2.52$ & -- & -- & -- & -- &   NM\\
DES\,J033556.28$-$540316.3 & 53.98449 & -54.05452 & 18.85 & 18.37 & $65.15 \pm 1.18$ & $1.56 \pm 0.24$ & $-2.88 \pm 0.18$ & $1.24 \pm 0.28$ & $-3.14 \pm 0.27$ &    M\\
DES\,J033558.15$-$540204.8 & 53.99228 & -54.03466 & 19.30 & 18.82 & $65.67 \pm 1.12$ & $1.92 \pm 0.38$ & $-2.57 \pm 0.24$ & $1.33 \pm 0.48$ & $-2.97 \pm 0.40$ &    M\\
DES\,J033601.76$-$540405.5 & 54.00734 & -54.06819 & 19.56 & 19.06 & $63.25 \pm 1.40$ & $1.08 \pm 0.32$ & $-3.16 \pm 0.37$ & -- & -- &    M\\
DES\,J033606.25$-$540144.5 & 54.02604 & -54.02903 & 20.39 & 19.93 & $29.99 \pm 4.19$ & -- & -- & -- & -- &   NM\\
DES\,J033607.75$-$540235.6 & 54.03230 & -54.04321 & 17.43 & 16.82 & $59.07 \pm 0.92$ & $2.22 \pm 0.21$ & $-2.78 \pm 0.12$ & -- & -- &    M\\
DES\,J033618.68$-$535745.1 & 54.07784 & -53.96254 & 18.04 & 18.20 & $55.71 \pm 3.23$ & -- & -- & -- & -- &    M\\
DES\,J033621.86$-$540040.7 & 54.09109 & -54.01130 & 20.30 & 19.87 & $64.66 \pm 1.75$ & -- & -- & -- & -- &    M\\
DES\,J033622.83$-$535955.5 & 54.09513 & -53.99876 & 18.58 & 18.12 & $126.61 \pm 0.94$ & $5.25 \pm 0.33$ & -- & -- & -- &   NM\\
DES\,J033635.75$-$540156.9 & 54.14895 & -54.03248 & 18.05 & 17.61 & $131.78 \pm 0.94$ & -- & -- & -- & -- &   NM\\
DES\,J033635.78$-$540120.2 & 54.14909 & -54.02228 & 20.17 & 19.73 & $56.68 \pm 1.91$ & -- & -- & -- & -- &   M?\\
\\ \multicolumn{11}{c}{ \ruleline{GIRAFFE} } \\
DES\,J033523.85$-$540407.5 & 53.84938 & -54.06875 & 16.45 & 15.65 & $65.69$ & $2.42 \pm 0.19$ & $-2.90 \pm 0.10$ & -- & -- &    M\\
DES\,J033524.96$-$540230.7 & 53.85400 & -54.04186 & 19.69 & 19.17 & $67.99$ & $5.30 \pm 0.57$ & -- & -- & -- &   NM\\
DES\,J033539.04$-$535556.4 & 53.91267 & -53.93234 & 17.85 & 17.16 & $-20.65$ & $6.97 \pm 0.26$ & -- & -- & -- &   NM\\
DES\,J033548.04$-$540349.8 & 53.95017 & -54.06384 & 18.27 & 17.69 & $62.78$ & $3.45 \pm 0.22$ & $-2.02 \pm 0.11$ & -- & -- &   M?\\
DES\,J033603.90$-$540026.5 & 54.01625 & -54.00737 & 17.79 & 17.09 & $-25.16$ & $5.87 \pm 0.23$ & -- & -- & -- &   NM\\
DES\,J033623.52$-$540518.9 & 54.09798 & -54.08857 & 19.75 & 19.23 & $213.51$ & $4.33 \pm 0.55$ & -- & -- & -- &   NM\\
\\ \multicolumn{11}{c}{ \ruleline{GMOS} } \\
DES\,J033539.85$-$540458.1 & 53.91605 & -54.08281 & 18.33 & 18.59 & $69.30$ & -- & -- & -- & -- &    M\\

\enddata
\tablenotetext{a}{Quoted magintudes represent the PSF magnitude derived from the DES coadded images by SExtractor. The one exception is DES\,J033523.85$-$540407.5 which uses the median AUTO magnitude derived from the individual DES single epoch images.}
\end{deluxetable*}

\acknowledgements{This publication is based upon work supported by the
  National Science Foundation under grant AST-1108811.  We thank Dan
  Kelson for helpful conversations, Anna Frebel for providing the MIKE
  spectrum of HD~122563, and Becky Canning and Jimmy for helpful
  conversations on the reduction of VLT spectra.  We also thank the
  anonymous referee for suggestions that improved the presentation of
  the paper.  ACR acknowledges financial support provided by the
  PAPDRJ CAPES/FAPERJ Fellowship.  FS acknowledges financial support
  provided by CAPES under contract No.\ 3171-13-2.  This research has
  made use of NASA's Astrophysics Data System Bibliographic Services.

  M2FS was built through partial support via NSF/MRI grant AST-0923160
  to Mario Mateo, Ian Thompson and Steven Shectman as co-PIs and its
  construction was assisted by Jeffrey Crane and John I. Bailey~III.
  The M2FS observations used in this paper were carried out on our
  behalf by Mario Mateo and Matthew Walker, who were supported in part
  by NSF grants AST-1312967 (MM) and AST-1313045 (MW).  Additional
  partial operational support of M2FS has been provided by the
  Department of Astronomy of the University of Michigan and by
  Carnegie Observatories.

  Funding for the DES Projects has been provided by the
  U.S. Department of Energy, the U.S. National Science Foundation, the
  Ministry of Science and Education of Spain, the Science and
  Technology Facilities Council of the United Kingdom, the Higher
  Education Funding Council for England, the National Center for
  Supercomputing Applications at the University of Illinois at
  Urbana-Champaign, the Kavli Institute of Cosmological Physics at the
  University of Chicago, the Center for Cosmology and Astro-Particle
  Physics at the Ohio State University, the Mitchell Institute for
  Fundamental Physics and Astronomy at Texas A\&M University,
  Financiadora de Estudos e Projetos, Funda{\c c}{\~a}o Carlos Chagas
  Filho de Amparo {\`a} Pesquisa do Estado do Rio de Janeiro, Conselho
  Nacional de Desenvolvimento Cient{\'i}fico e Tecnol{\'o}gico and the
  Minist{\'e}rio da Ci{\^e}ncia, Tecnologia e Inova{\c c}{\~a}o, the
  Deutsche Forschungsgemeinschaft and the Collaborating Institutions
  in the Dark Energy Survey.  The DES data management system is
  supported by the National Science Foundation under Grant Number
  AST-1138766.  The DES participants from Spanish institutions are
  partially supported by MINECO under grants AYA2012-39559,
  ESP2013-48274, FPA2013-47986, and Centro de Excelencia Severo Ochoa
  SEV-2012-0234, some of which include ERDF funds from the European
  Union.
   
  The Collaborating Institutions are Argonne National Laboratory, the
  University of California at Santa Cruz, the University of Cambridge,
  Centro de Investigaciones En{\'e}rgeticas, Medioambientales y
  Tecnol{\'o}gicas-Madrid, the University of Chicago, University College
  London, the DES-Brazil Consortium, the University of Edinburgh, the
  Eidgen{\"o}ssische Technische Hochschule (ETH) Z{\"u}rich, Fermi
  National Accelerator Laboratory, the University of Illinois at
  Urbana-Champaign, the Institut de Ci{\`e}ncies de l'Espai (IEEC/CSIC),
  the Institut de F{\'i}sica d'Altes Energies, Lawrence Berkeley
  National Laboratory, the Ludwig-Maximilians Universit{\"a}t
  M{\"u}nchen and the associated Excellence Cluster Universe, the
  University of Michigan, the National Optical Astronomy Observatory,
  the University of Nottingham, The Ohio State University, the
  University of Pennsylvania, the University of Portsmouth, SLAC
  National Accelerator Laboratory, Stanford University, the University
  of Sussex, and Texas A\&M University.
}
{\it Facilities:} \facility{Magellan:Clay (M2FS)} \facility{VLT:Kueyen
  (GIRAFFE)} \facility{Gemini:South (GMOS)}

%{\it Facilities:} 
%\facility{This paper includes data gathered with the 6.5 meter
%  Magellan Telescopes located at Las Campanas Observatory, Chile.}
%\facility{Based in part on data obtained from the ESO Science
%  Archive Facility under request number 157689.}
%\facility{Based in part on observations obtained at the Gemini
%  Observatory, which is operated by the Association of Universities
%  for Research in Astronomy, Inc., under a cooperative agreement with
%  the NSF on behalf of the Gemini partnership: the National Science
%  Foundation (United States), the National Research Council (Canada),
%  CONICYT (Chile), the Australian Research Council (Australia),
%  Minist\'{e}rio da Ci\^{e}ncia, Tecnologia e Inova\c{c}\~{a}o
%  (Brazil) and Ministerio de Ciencia, Tecnolog\'{i}a e Innovaci\'{o}n
%  Productiva (Argentina).}

\bibliographystyle{apj}
\bibliography{main}

\end{document}